\documentclass[prb,aps,twocolumn,scratcl,superscriptaddress]{revtex4-1}

\usepackage{graphicx}
\usepackage{savesym}
\usepackage{amsfonts}
\usepackage{amsmath}
\usepackage{bm}
\usepackage{color}
\usepackage{subfigure}
\usepackage{multirow}

\usepackage[colorlinks,bookmarks=false,citecolor=blue,linkcolor=red,urlcolor=blue]{hyperref}

\begin{document}
\title{Bosonic Many-body Theory of Quantum Spin Ice}
\author{Zhihao Hao}
\affiliation{Department of Physics and Astronomy, University of Waterloo, Ontario, N2L 3G1, Canada} 
\author{Alexandre G. R. Day}
\affiliation{Department of Physics and Astronomy, University of Waterloo, Ontario, N2L 3G1, Canada}
\author{Michel J. P. Gingras}
\affiliation{Department of Physics and Astronomy, University of Waterloo, Ontario, N2L 3G1, Canada} 
\affiliation{Perimeter Institute for Theoretical Physics, Waterloo, Ontario, N2L 2Y5, Canada} 
\affiliation{Canadian Institute for Advanced Research, 180 Dundas Street West, Suite 1400, Toronto, ON, M5G 1Z8, Canada}
\date{\today}
\begin{abstract}
We carry out an analytical study of quantum spin ice, a U$(1)$ quantum spin liquid close to the classical spin ice solution for an effective spin $1/2$ model with anisotropic exchange couplings $J_{zz}$, $J_{\pm}$ and $J_{z\pm}$ on the pyrochlore lattice.  Starting from the quantum rotor model introduced by Savary and Balents in Phys. Rev. Lett. \textbf{108}, 037202 (2012), we retain the dynamics of both the spinons and gauge field sectors in our treatment. The spinons are described by a bosonic representation of quantum XY rotors while the dynamics of the gauge field is captured by a phenomenological Hamiltonian. By calculating the one-loop spinon self-energy, which is proportional to $J_{z\pm}^2$, we determine the stability region of the U$(1)$ quantum spin liquid phase in the $J_{\pm}/J_{zz}$ vs $J_{z\pm}/J_{zz}$ zero temperature phase diagram. From these results, we estimate the location of the boundaries between the spin liquid phase and classical long-range ordered phases. 
\end{abstract}
\maketitle
\section{Introduction}
The search for quantum spin liquid states \cite{Balents.2010} has captured the interest of condensed matter physicists since the pioneering work of Anderson \cite{Anderson.1973}. Once proposed \cite{Anderson.1987} as a crucial element for the physics of copper-based high-temperature superconductors, quantum spin liquids now form an independent subfield on their own merit. In part, this is because they are predicted to possess intriguing properties such as fractionalized excitations and topological order \cite{Balents.2010}. Conceptually novel, such traits are also attractive for potential applications in quantum computation and quantum information processing \cite{Nayak.2008}.

\begin{figure}
\centering
\includegraphics[width=0.95\columnwidth]{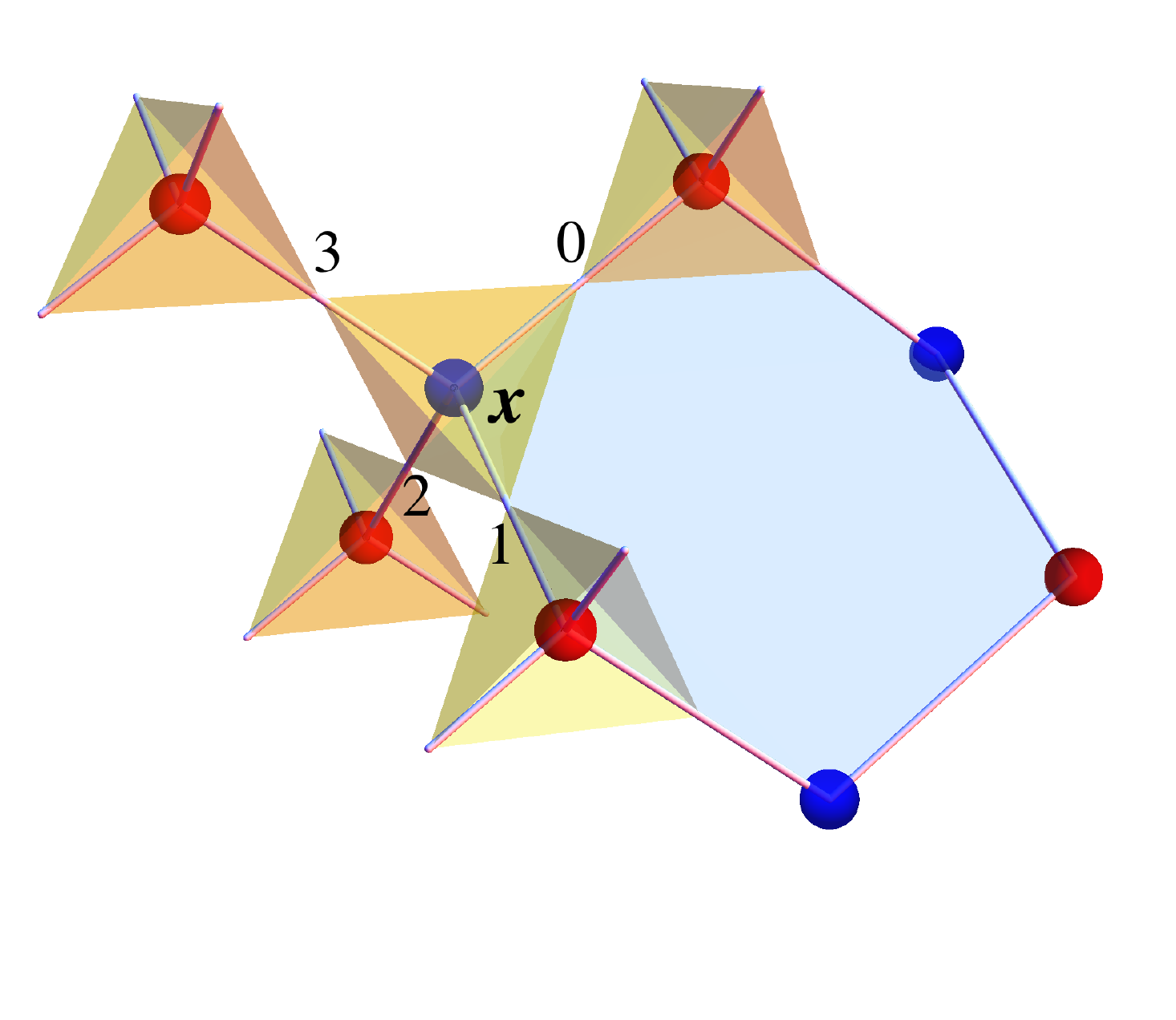}
\caption{The pyrochlore lattice and its medial lattice, the diamond lattice. The red and blue spheres are the A and B sublattices of the diamond lattice, respectively. $\boldsymbol{x}$ labels sites on the diamond lattice. $\alpha=0,1,2,3$ labels the four sites of the primitive basis  of the pyrochlore lattice. $\hat{\mu}$ denotes the four vectors connecting site $\boldsymbol{x}$ on the A sublattice to its nearest neighbors on the diamond lattice. The light-blue shaded region highlights a hexagonal plaquette on the diamond lattice. The lowest order quantum tunnelling process between two spin ice configurations involves the flip of six alternating spins around the plaquette. }\label{fig:pyro}
\end{figure}

Recently, a new avenue toward the discovery of quantum spin liquid phases has been uncovered in the form of highly anisotropic spin models on the pyrochlore lattice \cite{Molavian.2007,Onoda.2010,Ross.2011,Savary.2012,Lee.2012,Huang.2014,Gingras.2014}, a three-dimensional network of corner sharing tetrahedra (Figure \ref{fig:pyro}). The initial insight in the physics of these systems starts with Anderson's realization \cite{anderson.1956} that the antiferromagnetic Ising model on the pyrochlore lattice has an extensive number of ground states. For any of the ground states, there are two up spins and two down spins per tetrahedron \footnote{Ising spins along the global $z$ directions on the pyrochlore lattice are \emph{not} related to each other by point group operations of the octahedral group, O$_h$. Consequently, the antiferromagnetic Ising model with global Ising spins is not allowed by lattice symmetries for the pyrochlore lattice. On the other hand, local Ising spins as discussed in the main text transform into each other under the O$_h$ point group symmetries. }. These spin orientations are an exact mapping\footnote{Strictly speaking, common water ice is the hexagonal ice, $I_h$. Spin ice is really the exact mapping of the cubic ice (Ice VII) phase \cite{BramwellandGingras.2001, LMMmichelchapter}. } of the proton disordered configurations in water ice \cite{Pauling.1935} where each oxygen ion forms two strong and two weak hydrogen bonds with four protons, the so-called ``ice-rule'' \cite{Bernal.1933}.

Interest in the magnetic version of water ice,``spin ice'' \cite{BramwellandGingras.2001,LMMmichelchapter,Gardner_RMP} eventually intensified thanks to the discovery of two materials, Ho$_2$Ti$_2$O$_7$ \cite{Harris.1997,Harris.1998,Bramwell.2001} and Dy$_2$Ti$_2$O$_7$ \cite{Ramirez.1999}  embedding such physics. In both compounds, the combination of spin-orbit coupling and crystal-field effects mandates the magnetic moment of the rare earth ions, Ho$^{3+}$ and Dy$^{3+}$, to strictly point along the local $\langle 111\rangle$ directions \cite{ LMMmichelchapter,Gardner_RMP}. In the above two materials, the interactions between these local Ising spins are effectively antiferromagnetic because of the dipolar interaction \cite{Hertog.2000,Gingras.2001,Isakov.2005}. Treating the spins as ``magnetic fluxes'' of an emergent U$(1)$ gauge field \cite{Isakov.2004,Henley.2005,Henley.2010}, the low temperature spin-spin correlations of a spin ice system are well described by divergent-free spin configurations \cite{Fennell.2009,Morris.2009,Kadowaki.2009}, a direct translation of the ice-rule. Moreover, the low temperature properties of these materials are also well accounted by a low concentration of charges for the gauge field \cite{Fennell.2009,Morris.2009,Kadowaki.2009}, referred to as ``magnetic monopoles'' in Ref.~[\onlinecite{Castelnovo.2008}]. In this work, we adopt a dual perspective in which the local $[111]$ Ising component of the spin moment is mapped to electric flux of a gauge theory and the charge particles are spinons carrying ``electric charge'' \cite{Hermele.2004,Savary.2012}. 

The equilibrium thermodynamic properties of classical spin ice is now well understood \cite{ LMMmichelchapter}. As discussed above, the low energy divergent-free spin states are mapped to electric field configurations with no charges present and the low energy gapped excitations \footnote{Starting from a divergent-free spin-ice state, flipping a spin creates two charged spinons costing a finite amount of exchange energy. These excitations are thus gapped.} are spinons carrying an ``electric'' charge. However, inherent to such a classical system, neither the electric field nor the spinons have a dynamics. Quantum fluctuations of the spins are expected to endow the spinons and the gauge field with quantum dynamics. It is thus interesting to study theoretically the effect of quantum fluctuations in spin ice. Moreover, such an investigation is motivated by the exotic properties displayed by several materials in the same family as Ho$_2$Ti$_2$O$_7$ and Dy$_2$Ti$_2$O$_7$ including Yb$_2$Ti$_2$O$_7$ \cite{Hodges.2002,Ross.2011}, Tb$_2$Ti$_2$O$_7$ \cite{Gardner.1999},  Pr$_2$Sn$_2$O$_7$ \cite{Zhou.2008} and Pr$_2$Zr$_2$O$_7$ \cite{Kimura.2013}. All these materials are believed to have substantial interactions among all three moments of the effective spin $1/2$ moment  \cite{Molavian.2007,Onoda.2010,Ross.2011,Applegate.2012,Hayre.2013} in addition to the interaction between the Ising components, as in classical spin ice.

The theoretical investigation of quantum fluctuations in spin ice was engendered by Hermele \emph{et al.}'s work \cite{Hermele.2004}. Starting from an XXZ model on the pyrochlore lattice in the easy-axis anisotropy (Ising) limit, the authors of Ref.~[\onlinecite{Hermele.2004}] used degenerate perturbation theory to construct a low-energy effective theory of the XXZ model that incorporates the lowest order quantum tunnelling process between two classical spin-ice configurations. The resulting multi-spin motion flips six alternating spins around a hexagonal plaquette on the diamond lattice formed by the centers of tetrahedron (Fig.~\ref{fig:pyro}) \footnote{Interestingly, such spin-flip loops, or ``worms'', are the very low-energy excitations that facilitates effective simulations \cite{Melko.2004} to find the ground state of the classical dipolar spin ice model \cite{Hertog.2000}.}. The Ising components of one direction, $+1$ for example, can be mapped as hard core dimers living on the bonds of the diamond lattice.  By leveraging the extensive knowledge of the properties of the quantum dimer model \cite{Gingras.2014,Huse.2003,Moessner.2003, LMMqdmchapter}, the effective theory can be described by a dynamical compact U$(1)$ gauge theory in its deconfined phase. Both the quantum ground state and its gapless gauge fluctuations are coherent superpositions of classical spin-ice configurations. The predicted U$(1)$ liquid was later found in quantum Monte-Carlo studies \cite{Banerjee.2008,Kato.2014} of the XXZ model at finite temperature. The properties of the spin liquid have been further characterized in detail by both analytical calculations and quantum Monte-Carlo simulations \cite{Shannon.2012,Benton.2012} of the dimer model at $T=0$. 

These works \cite{Hermele.2004,Shannon.2012,Benton.2012} focus on the ground state of the quantum spin ice and the ``photon'' excitations, charge-neutral gauge fluctuations with respect to the ground state. The first study of ``electrically charged matter'', spinons excitations, in quantum spin ice \footnote{There have, however, been prior studies of spinon excitations on the three-dimensional pyrochlore lattice in other contexts than the quantum spin ice problem studied here. For example, see Ref. \cite{Nussinov.2007}. } was performed by Savary and Balents \cite{Savary.2012}. Starting from a symmetry-motivated anisotropic spin Hamiltonian \cite{Ross.2011,Savary.2012} on the pyrochlore lattice \footnote{Previous works had investigated the role of anisotropic spin couplings on the pyrochlore lattice \cite{Molavian.2007,Curnoe.2004,McClarty.2009,Cao.2009,Onoda.2010,Thompson.2011}.}, they introduced a mapping where the spinons are represented by quantum XY rotors. These rotors interact with the dynamical compact U$(1)$ gauge field discussed in the previous paragraph. The authors solved their model using a gauge mean-field approximation (g-MFT) \cite{Savary.2012}, neglecting the dynamics and correlations of the gauge fields. They established a phase diagram with two quantum spin liquid states: the aforementioned U$(1)$ spin liquid \cite{Hermele.2004} with all components of the spins disordered quantum mechanically and a novel ``hybrid'' state they named Coulomb ferromagnet (CFM). According to the g-MFT calculations of Ref.~[\onlinecite{Savary.2012}], the CFM phase is characterized by ordered Ising components of the spins coexisting with deconfined spinons. However, we note in Appendix \ref{app:5} that the XY components of the spins also have static expectation values within the g-MFT formalism. The coexistence of full long range magnetic order in both the Ising components and XY comments with deconfined spinons instead of conventional magnons highlights the peculiar and yet intriguing property of the CFM phase. The same g-MFT approach was later applied to systems with non-Kramers magnetic ions to propose the possibility of a Z$_2$ spin liquid \cite{Lee.2012}. In a separate and more recent development, Huang \emph{et al}. \cite{Huang.2014} identified a XYZ model as the minimum description of materials where the spin wave functions are linear combinations of $|J_{z}\rangle$ in the local frame with $J_{z}=3n/2$, $n$ being an odd integer. Reference~[\onlinecite{Huang.2014}] used g-MFT to analyze their model and proposed U$(1)$ and Z$_2$ spin liquids as two possible quantum spin liquids of the XYZ model. 

While these exciting developments are contributing to our understanding of possible quantum spin liquid phases in the vicinity of the classical spin ice solution, some important questions remain. Physically, all analytical approaches \cite{Hermele.2004,Benton.2012,Savary.2012,Lee.2012,Huang.2014,Castro.2006} describe the dynamics of either the neutral gauge fluctuations \emph{or} the charged spinons. In practice, \emph{both} types of excitations have their own dynamics while \emph{interacting} non-trivially. It is thus desirable to perform a study of quantum spin ice with both types of excitations considered dynamical. Moreover, for the description of spinons,  the solution of the quantum XY rotor model relies on the ``large-$N$'' approximation while $N=1$ for XY rotors. The large-$N$ approach is not straightforwardly amenable to improvement via standard diagrammatic many-body treatments. It is therefore desirable to explore alternative schemes for which conventional many-body techniques and approximation schemes can be readily applied.  

With these motivations laid out, we present here a study of the anisotropic spin model of Eq.~\eqref{eqn:hami} investigated in Ref.~[\onlinecite{Savary.2012}] with both gauge fluctuation and spinon dynamics now included. Starting from the quantum rotor model \cite{Savary.2012}, we introduce the dynamics of the gauge field and separate the gauge field into a static part and a fluctuating part. Under the background of the static part of the gauge field, we study the physics of the spinon sector by introducing a bosonic representation of the XY rotors. We find that both the ground state and a single spinon energy reduce to the expected forms in the classical limit. The gauge fluctuations are included by borrowing from the work of Benton \emph{et al}. \cite{Benton.2012}. To estimate the effect of the quantum fluctuations introduced by the interaction between the Ising and XY components of the spin, $J_{z\pm}$, we calculate the one-loop correction to the spinon self-energy to second order in $J_{z\pm}$. By identifying the boundary for spinon condensation, we establish the perturbatively stable region of the U$(1)$ liquid phase. Using a combination of energy calculations and numerical results from a previous study \cite{Isakov.2004}, we construct a zero-temperature phase diagram in the anisotropic exchange constants. 

The rest of paper is as follows. We present in Section \ref{sec:intro} the model and separate it into a spinon sector, a gauge fluctuation sector and the interaction between spinons and the gauge field. We study the dynamics of the spinons and the gauge field in Section \ref{sec:spinon} and \ref{sec:gfield}, respectively. The energy of the spinon modes are modified by the self-energy contribution proportional to $J_{z\pm}^2$. For $J_{z\pm}$ beyond some critical threshold value, the spinon mode at zero momentum condenses and the U$(1)$ liquid is destroyed, giving way to either a splayed ferromagnetic (SFM) phase \cite{Ross.2011,Yaouanc.2013,Wong.2013} or an XY antiferromagnetic phase \cite{Champion.2003,Wong.2013}. We present the calculation of the spinon self-energy as well as  the phase diagram in Section \ref{sec:senergy}. We conclude our paper in Section \ref{sec:dis} by discussing connections of our work with previous studies and identifying directions for future studies. The reader is provided with a number of appendices for the technical details of the calculations. 

\section{General formalism}\label{sec:intro}
To simplify the notation in what follows, the pyrochlore lattice is represented using its medial lattice, the diamond lattice (Fig.~\ref{fig:pyro}). The sites of the diamond lattice are labeled by $\boldsymbol{x}$. Each diamond lattice site is connected to four nearest-neighbor sites by vectors $\hat{\mu}$ ($\mu=0,1,2,3$).  The explicit expressions for the (non-unit) vectors $\hat{\mu}$ are given in Appendix \ref{app:1}. Each spin resides at the middle point of the diamond lattice bonds. Using the fact that the diamond lattice can be separated into two interpenetrating face-centered cubic sublattices, labelled as A and B sublattices, a bond connecting the A sublattice site $\boldsymbol{x}$ and B sublattice site $\boldsymbol{x}+\hat{\mu}$ is labeled as $\langle\boldsymbol{x}\mu\rangle$. A spin at the center of the bond $\langle\boldsymbol{x}\mu\rangle$ is written as $\boldsymbol{S}_{\boldsymbol{x}\mu}$. With these notations in place, the spin Hamiltonian \cite{Savary.2012} that we study reads in terms of the local spin components:
\begin{subequations}\label{eqn:hami}
\begin{eqnarray}
&&\mathcal{H}=\sum_{\boldsymbol{x}}\frac{J_{zz}}{2}Q_{\boldsymbol{x}}^2-\label{eqn:h0}\\&&\sum_{\boldsymbol{x}\in\langle A\rangle}\left[J_{\pm}\sum_{\mu<\nu}(S_{\boldsymbol{x}\mu}^+S_{\boldsymbol{x}\nu}^-+S_{\boldsymbol{x}\mu}^+S_{\boldsymbol{x}+\hat{\mu}-\hat{\nu},\nu}^-+h.c)\right.\label{eqn:hpm}\\&&\left.-J_{z\pm}\sum_{\mu\neq\nu}\left(S_{\boldsymbol{x}\mu}^z(S_{\boldsymbol{x}\nu}^++S_{\boldsymbol{x}+\hat{\mu}-\hat{\nu},\nu}^+)\mathrm{e}^{i\gamma_{\mu\nu}}+h.c\right)\right]
\end{eqnarray}
\end{subequations}
where $\gamma_{01}=\gamma_{23}=0$, $\gamma_{02}=\gamma_{13}=-2\pi/3$ and $\gamma_{03}=\gamma_{12}=2\pi/3$ (Refs.~[\onlinecite{Ross.2011,Savary.2012}]). $\langle A\rangle$ and $\langle B\rangle$ denote the collection of sites on the A and B sublattices, respectively. The charge $Q_{\boldsymbol{x}}$ is related to $S^{z}_{\boldsymbol{x}\mu}$ by:
\begin{equation}\label{eqn:charged}
Q_{\boldsymbol{x}}=\left\{\begin{array}{lll}\sum_{\mu}S^{z}_{\boldsymbol{x}\mu} &\mathrm{for} &\boldsymbol{x}\in\langle A\rangle,\\ -\sum_{\mu}S^{z}_{\boldsymbol{x}-\hat{\mu},\mu} &\mathrm{for}&\boldsymbol{x}\in \langle B\rangle \end{array}\right.
\end{equation}
We note that  Eq.~\eqref{eqn:hami} is not the most general nearest-neighbor Hamiltonian on the pyrochlore lattice since the interaction between $S_{\boldsymbol{x}\mu}^{+}$ and $S_{\boldsymbol{x}\nu}^{+}$ (Ref.~[\onlinecite{Ross.2011,Savary.2012,Lee.2012,Huang.2014}]), with coupling $J_{\pm\pm}$, is omitted. Our main goal in this paper is  to explore a many-body formulation of quantum spin ice in a simple yet non-trivial context. The anisotropic $J_{\pm\pm}$ exchange coupling leads to four spinon interaction \cite{Lee.2012,Huang.2014} whose treatment is beyond such a scope. The study of its effect will be left for the future work. 

Following Ref.~[\onlinecite{Savary.2012}], we introduce a rotor representation of Eq.~\eqref{eqn:hami}. On each site of the diamond lattice, a pair of conjugate operators, $\hat{Q}_{\boldsymbol{x}}$ and $\hat{\theta}_{\boldsymbol{x}}$, are introduced satisfying the following commutation relation:
\begin{equation}\label{eqn:commut}
[\hat{\theta}_{\boldsymbol{x}},\hat{Q}_{\boldsymbol{x}^\prime}]=i\delta_{\boldsymbol{x}\boldsymbol{x}^\prime}.
\end{equation}
Starting from Eq.~\eqref{eqn:commut} and using $\psi_{\boldsymbol{x}}\equiv \mathrm{e}^{-i\hat{\theta}_{\boldsymbol{x}}}$, the following relation can be derived:
\begin{equation}\label{eqn:com2}
[\psi_{\boldsymbol{x}},\hat{Q}_{\boldsymbol{x}^\prime}]=\psi_{\boldsymbol{x}}\delta_{\boldsymbol{x}\boldsymbol{x}^\prime}.
\end{equation}
Taking integer eigenvalues, $\hat{Q}_{\boldsymbol{x}}$ represents the charge on site $\boldsymbol{x}$. $\psi_{\boldsymbol{x}}$ decreases the charge on site $\boldsymbol{x}$ by one while $\psi_{\boldsymbol{x}}^\dagger$ increases it by one. 


The transverse components of the spins, $S_{\boldsymbol{x}\mu}^{\pm}$, are represented using the rotor operators in addition with pseudo-spin operators $s_{\boldsymbol{x}\mu}^{\pm}$ \cite{Savary.2012}:
\begin{equation}\label{eqn:transverse}
S_{\boldsymbol{x}\mu}^{+}=\psi_{\boldsymbol{x}}^\dagger s^{+}_{\boldsymbol{x}\mu}\psi_{\boldsymbol{x}+\hat{\mu}},\qquad S_{\boldsymbol{x}\mu}^-=\psi_{\boldsymbol{x}+\hat{\mu}}^\dagger s^{-}_{\boldsymbol{x}\mu}\psi_{\boldsymbol{x}}. 
\end{equation}
In Hamiltonian Eq.~\eqref{Savary.2012}, we have shifted the energy by $J_{zz}/4$ per spin so that the energy of the classical spin ice state is zero. For the Ising component of the pseudo-spins $\boldsymbol{s}_{\boldsymbol{x}\mu}$, we have $s_{\boldsymbol{x}\mu}^z\equiv S_{\boldsymbol{x}\mu}^{z}$. We henceforth omit distinguishing $s_{\boldsymbol{x}\mu}^z$ and $S_{\boldsymbol{x}\mu}^{z}$. The mapping in Eq.~\eqref{eqn:transverse} preserves the correct commutation relations of the spin components of the physical spin $\boldsymbol{S}_{\boldsymbol{x}\mu}$ in ${\cal H}$. The transverse components of the pseudo-spin is mapped to an exponential function of the vector gauge field, $A_{\boldsymbol{x}\mu}$: 
\begin{equation}\label{eqn:gaugefield}
s^{\pm}_{\boldsymbol{x}\mu}\to \frac{1}{2}\mathrm{e}^{\pm iA_{\boldsymbol{x}\mu}}.
\end{equation}
$S_{\boldsymbol{x}\mu}^{z}$ is then represented as the  electric flux of the gauge field: $S_{\boldsymbol{x}\mu}^{z}\to E_{\boldsymbol{x}\mu}$.  We note that the mapping of pseudo-spin operators in terms of gauge field operators $A_{\boldsymbol{x}\mu}$ and $E_{\boldsymbol{x}\mu}$ adopted here is inspired by their expectation values in a spin coherent state $|\cos\theta,\phi\rangle$ where $\theta$ and $\phi$ are polar and azimuthal angles, respectively. The $1/2$ prefactor in Eq.~\eqref{eqn:gaugefield} is just the spin length. The electric flux, $E_{\boldsymbol{x}\mu}$, and the gauge field, $A_{\boldsymbol{x}\mu}$, satisfy the following commutation relation:
\begin{equation}\label{eqn:comm2}
[A_{\boldsymbol{x}\mu},E_{\boldsymbol{x}^\prime\nu}]=i\delta_{\boldsymbol{x}\boldsymbol{x}^\prime}\delta_{\mu\nu}.
\end{equation}
While Eq.~\eqref{eqn:comm2} is expected from the standard Hamiltonian formulation of quantum electrodynamics \cite{Kogut.1979}, it also captures the physics that $s^{\pm}_{\boldsymbol{x}\mu}$, proportional to $\exp(\pm iA_{\boldsymbol{x}\mu})$, increases or decreases $S^{z}_{\boldsymbol{x}\mu}$, directly translated into $E_{\boldsymbol{x}\mu}$, by 1. Physically, $S^{+}_{\boldsymbol{x}\mu}$ creates a positive spinon at site $\boldsymbol{x}$ and a negative spinon at site $\boldsymbol{x}+\hat{\mu}$. $\mathrm{e}^{iA_{\boldsymbol{x}\mu}}$ changes the electric flux on bond $\langle\boldsymbol{x}\mu\rangle$ so that the ``Gauss Law'' \eqref{eqn:charged} is preserved. 
The temporal components of the gauge field, which we define as $\phi_{\boldsymbol{x}}$, are Lagrange multipliers introduced to enforce Eq.~\eqref{eqn:charged}. Setting $J_{zz}=1$ as the overall energy scale, we write $j_{\pm}\equiv J_{\pm}/J_{zz}$ and $j_{z\pm}\equiv J_{z\pm}/J_{zz}$. Using the $A_{\boldsymbol{x}\mu}$, $E_{\boldsymbol{x}\mu}$ and $\psi_{\boldsymbol{x}}$ fields, the Hamiltonian \eqref{eqn:hami} is rewritten as:
\begin{widetext}
\begin{subequations}\label{eqn:hamigauge}
\begin{eqnarray}
&&\mathcal{H}=\frac{1}{2}\sum_{\boldsymbol{x}}\hat{Q}_{\boldsymbol{x}}^2-
\sum_{\boldsymbol{x}\in\langle A\rangle}\left[\frac{j_{\pm}}{4}\sum_{\mu<\nu}(\psi_{\boldsymbol{x}}^\dagger \mathrm{e}^{i(A_{\boldsymbol{x}\mu}-A_{\boldsymbol{x}+\hat{\mu}-\hat{\nu},\nu})}\psi_{\boldsymbol{x}+\hat{\mu}-\hat{\nu}}+\psi_{\boldsymbol{x}+\hat{\mu}}^\dagger \mathrm{e}^{-i(A_{\boldsymbol{x}\mu}-A_{\boldsymbol{x}\nu})}\psi_{\boldsymbol{x}+\hat{\nu}}+h.c)\right.\label{eqn:hh0}\\
&&\left.-\frac{j_{z\pm}}{2}\sum_{\mu\neq\nu}\left(E_{\boldsymbol{x}\mu}(\psi_{\boldsymbol{x}}^\dagger\mathrm{e}^{iA_{\boldsymbol{x}\nu}}\psi_{\boldsymbol{x}+\hat{\nu}}+\psi_{\boldsymbol{x}+\hat{\mu}-\hat{\nu}}^\dagger\mathrm{e}^{iA_{\boldsymbol{x}+\hat{\mu}-\hat{\nu},\nu}}\psi_{\boldsymbol{x}+\hat{\mu}})\mathrm{e}^{i\phi_{\mu\nu}}+h.c\right)\right]\label{eqn:int}\\ &&+\frac{U}{2}\sum_{\boldsymbol{x}\in \langle A\rangle,\mu}E_{\boldsymbol{x},\boldsymbol{x}+\hat{\mu}}^2+\sum_{\boldsymbol{x}}\phi_{\boldsymbol{x}}(\hat{Q}_{\boldsymbol{x}}-Q_{\boldsymbol{x}}). \label{eqn:addition}
\end{eqnarray}
\end{subequations}
\end{widetext}
Comparing with the Hamiltonian studied in Refs.~[\onlinecite{Savary.2012}], we have added Eq.~\eqref{eqn:addition}. The term proportional to $U$ controls the dynamics of gauge field as in standard quantum electrodynamics $\boldsymbol{E}\sim -\partial \boldsymbol{A}/\partial t$. In Ref.~[\onlinecite{Hermele.2004}] and [\onlinecite{Benton.2012}], this term was introduced to enforce the constrain $E_{\boldsymbol{x}\mu}=\pm 1/2$. This term can reproduce the dynamical structure factor computed using quantum Monte-Carlo simulations \cite{Benton.2012}. The second term in \eqref{eqn:addition} is the aforementioned Lagrange multiplier term enforcing the lattice Gauss Law. The Hamiltonian \eqref{eqn:hamigauge} represents the starting point of our work. The theory \eqref{eqn:hamigauge} is invariant under the following \emph{local} U$(1)$ gauge transformation: 
\begin{subequations}\label{eqn:gaugetransform}
\begin{eqnarray}
\psi_{\boldsymbol{x}}&\to& \psi_{\boldsymbol{x}}\mathrm{e}^{i\alpha_{\boldsymbol{x}}},\\
A_{\boldsymbol{x}\mu} &\to&A_{\boldsymbol{x}\mu}+\alpha_{\boldsymbol{x}}-\alpha_{\boldsymbol{x}+\hat{\mu}}.
\end{eqnarray}
\end{subequations}
From now on, we shall work in the temporal gauge where $\phi_{\boldsymbol{x}}\equiv 0$. 

To make progress, we separate the gauge field $A_{\boldsymbol{x}\mu}$ into a static part, $\bar{A}_{\boldsymbol{x}\mu}$, and a fluctuating part, $\tilde{A}_{\boldsymbol{x}\mu}$:
\begin{equation}
A_{\boldsymbol{x}\mu}=\bar{A}_{\boldsymbol{x}\mu}+\tilde{A}_{\boldsymbol{x}\mu}. 
\end{equation}
Following Refs.~[\onlinecite{Savary.2012}] and [\onlinecite{Hermele.2004}], we assume that the background gauge field $\bar{A}$ leads to zero magnetic fluxes through all hexagonal plaquettes. We are free to choose a gauge such that $\bar{A}_{\boldsymbol{x}\mu}=0$ for all bonds $\langle \boldsymbol{x}\mu\rangle$. To demonstrate that this is possible, we consider a pyrochlore lattice of $4V$ sites. There are $4V$ $\bar{A}_{\boldsymbol{x}\mu}$ fields. As for any vector fields, $\bar{A}_{\boldsymbol{x}\mu}$ can be separated into a longitudinal part and a transverse part. The transverse contributions are set to zero by the magnetic fluxes through $4V$ hexagonal plaquettes related to $\bar{A}_{\boldsymbol{x}\mu}$ by the lattice version of Stokes' theorem.  Out of the $4V$ equations, we note that there are only $2V$ linearly independent constraints. The corresponding $2V$ transverse components are zero since all magnetic fluxes are assumed to be zero.  The remaining $2V$ longitudinal components can be fixed to be zero by tuning $2V$ $\alpha_{\boldsymbol{x}}$ values.   We split the Hamiltonian \eqref{eqn:hamigauge} into three terms:
\begin{equation}\label{eqn:threeparts}
\mathcal{H}=H_{\mathrm{s}}(\hat{Q}_{\boldsymbol{x}},\psi_{\boldsymbol{x}})+H_{\mathrm{g}}(E_{\boldsymbol{x}\mu},\tilde{A}_{\boldsymbol{x}\mu})+H_{\mathrm{int}}(E_{\boldsymbol{x}\mu},\tilde{A}_{\boldsymbol{x}\mu},\psi_{\boldsymbol{x}}).
\end{equation}
The explicit forms of the spinon Hamiltonian, $H_{\mathrm{s}}$, the gauge field Hamiltonian, $H_{\mathrm{g}}$ and the interaction Hamiltonian, $H_{\mathrm{int}}$ are individually discussed in the following three sections. 

\section{Spinons}\label{sec:spinon}
In this section we focus on the spinon fields, $\psi_{\boldsymbol{x}}$. With a background of zero magnetic fluxes per hexagonal plaquette, the spinon Hamiltonian is:
\begin{equation}\label{eqn:hm}
\begin{aligned}
H_{\mathrm{s}}=&\frac{1}{2}\sum_{\boldsymbol{x}}\hat{Q}_{\boldsymbol{x}}^2-\frac{j_{\pm}}{4}\sum_{\boldsymbol{x}\in\langle A\rangle}\sum_{\mu<\nu}\left[\psi_{\boldsymbol{x}}^\dagger\psi_{\boldsymbol{x}+\hat{\mu}-\hat{\nu}}\right.\\&\left.+\psi_{\boldsymbol{x}+\hat{\mu}}^\dagger\psi_{\boldsymbol{x}+\hat{\nu}}+h.c\right]
\end{aligned}
\end{equation} 
In Eq.~\eqref{eqn:hm}, spinons of opposite charges can be created on neighboring sites on one diamond sublattice, sublattice A for example. Moreover, a spinon at a site on either the A or B sublattice can only hop to adjacent sites on the \emph{same} sublattice. The two sublattices support two independent but identical copies of the spinon Hamiltonian. Consequentially, we focus only on the dynamics within the A sublattice, a face-centered cubic (FCC) lattice which is the primitive space lattice of the diamond lattice. We note that under the fixed background gauge field, the constraints of the spinon motion, which has been shown to affect on the diffusive motion of spinons (a.k.a ``magnetic monopoles'' \cite{Castlenovo.2008}) in the classical spin ice \cite{Jaubert.2011}, are ignored here. For the FCC lattice, the spinon Hamiltonian $H_{\mathrm{s}}$ (Eq.~\eqref{eqn:threeparts}) is written as:
\begin{equation}\label{eqn:hma}
H_{\mathrm{s}}=\sum_{\boldsymbol{x}}\left[\frac{1}{2}\hat{Q}_{\boldsymbol{x}}^2-\frac{j_{\pm}}{4}\sum_{\mu<\nu}(\psi_{\boldsymbol{x}}^\dagger\psi_{\boldsymbol{x}+\hat{\mu}-\hat{\nu}}+h.c)\right].
\end{equation}

In previous works \cite{Savary.2012,Lee.2012,Huang.2014}, the rotor model \eqref{eqn:hma} was solved by relaxing the local constrain $|\psi_{\boldsymbol{x}}|=1$ to a global constrain $\sum_{\boldsymbol{x}}(|\psi_{\boldsymbol{x}}|^2-1)=0$ enforced by a Lagrange multiplier. The approximation can be regarded as a large-$N$ approximation for an O$(N)$ rotor where the local fluctuations of the $|\psi_{\boldsymbol{x}}|$ are suppressed by $1/N$. On the other hand,the XY rotor used here has $N=1$. The large-$N$ approach makes the theory amenable to an analytical treatment that leads to a qualitative insight on the possible phases of the rotor model. The corrections to the approximation can be calculated by accounting for $1/N$ contributions order-by-order. 

In this work, we adopt an alternative approximation scheme to that of Refs.~[\onlinecite{Savary.2012,Lee.2012,Huang.2014}].  Here, we introduce a bosonic representation of a quantum XY rotor similar to the well-known Holstein-Primakoff boson representation \cite{assa} of spins:
\begin{subequations}\label{eqn:hpboson}
\begin{eqnarray}
\psi&=&\frac{1}{\sqrt{1+d^\dagger d+b^\dagger b}}(d+b^\dagger),\\
\hat{Q}&=&d^\dagger d-b^\dagger b. 
\end{eqnarray}
\end{subequations} 
The $d$ and $b$ bosons carry positive and negative charge, respectively. To enforce the $|\psi|=1$ constraint, we demand that the two type of bosons cannot appear simultaneously. Defining $n_b\equiv b^{\dagger}b$ and $n_{d}\equiv d^\dagger d$, the constraint translates into:
\begin{equation}\label{eqn:cons}
n_bn_d=0.
\end{equation}
 As a result, products $bd$ and $b^\dagger d^\dagger$ are identically zero for all basis states satisfying the constraint. The two types of bosons are mutually exclusive: we thus name the representation ``exclusive bosons''.  This requirement can also be understood by examining the Hilbert space of a single quantum XY rotor. In the ``charge'' representation, the basis states are discrete states $|Q\rangle$ with $\hat{Q}|Q\rangle=Q|Q\rangle$. In the boson language, states with positive or negative charges are represented by $(d^+)^{Q}|0\rangle$ and $(b^{+})^{-Q}|0\rangle$ where $|0\rangle$ is the vacuum state with no bosons and zero charge. However, a rotor state with charge $Q$ has infinite more bosonic incarnations $(d^+)^{n}(b^{+})^m|0\rangle$ as long as $n-m=Q$ where $n$ and $m$ are both integers. By demanding that $b$ and $d$ bosons do not appear at the same time, i.e. $nm=0$, we recover a one-to-one mapping between the rotor and the boson Hilbert space. Formally, the exclusiveness, Eq.~\eqref{eqn:cons}, of the $b$ and $d$ bosons could be enforced by a large repulsion between the bosons, or by Lagrange multipliers. Under the representation \eqref{eqn:hpboson}, the commutation relation $[\psi,\hat{Q}]=\psi$ (Eq.~\eqref{eqn:com2}) is also preserved. We conclude that a pair of exclusive bosons is a faithful representation of a quantum XY rotor. 
 

We rewrite the spinon Hamiltonian \eqref{eqn:hma} using pairs of exclusive bosons defined separately on every site $\boldsymbol{x}$. We note that the exclusiveness applies only on-site: $d_{\boldsymbol{x}}$ boson and $b_{\boldsymbol{x}^\prime}$ boson do not appear simultaneously only if $\boldsymbol{x}=\boldsymbol{x}^\prime$. We normal-order $Q_{\boldsymbol{x}}^2$ with respect to the classical vacuum, or classical spin ice states, with no spinons:
\begin{equation}
Q_{\boldsymbol{x}}^2=d_{\boldsymbol{x}}^\dagger d_{\boldsymbol{x}}+b_{\boldsymbol{x}}^\dagger b_{\boldsymbol{x}}+d_{\boldsymbol{x}}^\dagger d_{\boldsymbol{x}}^\dagger d_{\boldsymbol{x}}d_{\boldsymbol{x}}+b_{\boldsymbol{x}}^\dagger b_{\boldsymbol{x}}^\dagger b_{\boldsymbol{x}}b_{\boldsymbol{x}}. 
\end{equation}
Assuming the boson densities are low for small $j_{\pm}$ in the quantum spin ice or U$(1)$ spin liquid, we neglect interactions among bosons as well as their exclusiveness. We assess below the validity of this approximation.  To keep the level of notation minimal, we hereafter use $\psi_{\boldsymbol{x}}$ to imply its lowest order (low density) bosonic approximation:
\begin{equation}\label{eqn:approxpsi}
\psi_{\boldsymbol{x}}\approx d_{\boldsymbol{x}}+b^\dagger_{\boldsymbol{x}}. 
\end{equation}
The original spinon Hamiltonian \eqref{eqn:hma} then becomes:
\begin{equation}\label{eqn:bosonq}
H\approx \sum_{\boldsymbol{x}}\left[\frac{1}{2}(d_{\boldsymbol{x}}^\dagger d_{\boldsymbol{x}}+b_{\boldsymbol{x}}^\dagger b_{\boldsymbol{x}})-\frac{j_{\pm}}{4}(\psi_{\boldsymbol{x}}^\dagger \psi_{\boldsymbol{x}+\hat{\mu}-\hat{\nu}}+h.c)\right]
\end{equation}

\begin{figure}
\centering
\includegraphics[width=0.95\columnwidth]{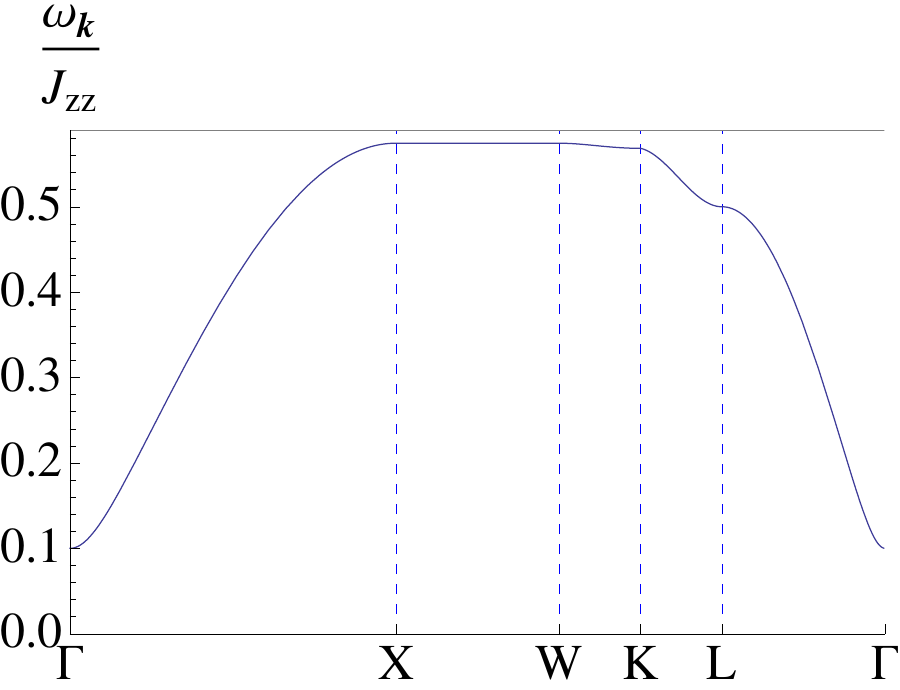}
\caption{The spinon dispersion $\omega_{\boldsymbol{k}}$ along high symmetry directions in the FCC Brillouin zone \cite{Kittel} for $j_{\pm}=0.08$.}\label{fig:spiondis}
\end{figure}

We write the bosons in terms of their Bloch modes and obtain the dispersion for the quasiparticles by a Bogoliubov transformation (see Appendix \ref{app:bogo}):
\begin{equation}\label{eqn:dispersion}
\omega_{\boldsymbol{k}}=\frac{1}{2}\sqrt{1-2 j_{\pm}\sum_{\alpha\neq\beta}\cos\frac{k_{\alpha}}{2}\cos\frac{k_{\beta}}{2}}
\end{equation}
where $\alpha,\beta=x,y,z$ are the three global cubic directions $[100]$, $[010]$ and $[001]$ (Fig. \ref{fig:spiondis}). The linear size $a_0\equiv1$ of the conventional cubic unit cell for the pyrochlore lattice is used as the elementary length unit. 

Considering first the limit of small $j_{\pm}$, the dispersion is then approximately:
\begin{equation}\label{eqn:app}
\omega_{\boldsymbol{k}}\approx \frac{1}{2}-\frac{j_{\pm}}{2}\sum_{\alpha\neq\beta}\cos\frac{k_{\alpha}}{2}\cos\frac{k_{\beta}}{2}. 
\end{equation}
We observe that in the limit of $j_{\pm}\to 0$, a single spinon cost energy $J_{zz}/2$, which agrees with the classical result \cite{ LMMmichelchapter}. 
Moreover, Eq.~\eqref{eqn:app} agrees with a simple variational estimate of the single-spinon dispersion using Hamiltonian \eqref{eqn:hma} without creating or annilating pairs of spinons (see Appendix \ref{app:4}).

From Eq.~\eqref{eqn:dispersion}, one finds that $\omega_{\boldsymbol{k}}$ vanishes at $\boldsymbol{k}=0$ for $j_{\pm}=1/12$: the spinons condense, leading to a Higgs phase. In terms of the physical spins, the state corresponds to long-range order of their transverse local $XY$ moments \cite{Wong.2013}. 

The ground state energy of the spinons per FCC unit cell is:
\begin{equation}\label{eqn:e0}
E_0=\frac{2}{V}\sum_{\boldsymbol{k}}\left(\omega_{\boldsymbol{k}}-\frac{1}{2}\right). 
\end{equation}
Here $V$ is the number of unit cells and the extra prefactor of $2$ comes from the two identical contributions from spinons on the A and B diamond sublattices. In the limit of small $j_{\pm}$, $E_0$ is found to be approximately given by:
\begin{equation}\label{eqn:e0app}
E_0\approx -\frac{3j_{\pm}^2}{2}-3j_{\pm}^3+O(j_{\pm}^4). 
\end{equation}
The ground state energy vanishes at $j_{\pm}=0$, agreeing with the energy of the ground state for  the spinon Hamiltonian \eqref{eqn:hm} in the same limit.   

We now proceed to check the internal consistency of our low boson density approximation by calculating the boson density $n\equiv\langle n_{b}+n_{d}\rangle$ (see Appendix \ref{app:bogo}). $n$ is a monotonic increasing function of $j_{\pm}$ which reaches its maximum value of approximately $0.029$ when the boson condenses at $j_{\pm}=1/12$, more than a factor $25$ smaller than the density of bosons, $n=3/4$,  in the high-temperature paramagnetic phase of the classical spin ice \footnote{At very high temperature, all spin configurations are equivalent. For a single tetrahedron, there are $2^4=16$ spin configurations. There are two ``doubly-charged'' configurations, all-in or all-out, with $2$ bosons. There are $8$ ``3-in-1-out'' configurations with one boson. The average number of bosons is thus $(2\times2+8)/16=3/4$.}.  We thus conclude that the dilute approximation is reasonable for $0\le j_{\pm}\le 1/12$. 


Once interactions induced by $H_{\mathrm{int}}\sim j_{z\pm}$ in Eq.~\eqref{eqn:threeparts} between spinons and gauge fields are taken into account, the spinon dispersion \eqref{eqn:dispersion} gets corrected by a self-energy contribution.In particular, the $j_{z\pm}$ interaction between the Ising and XY components of spins couples electric field $E_{\boldsymbol{x}\mu}$ and two powers of the spinon fields together. For $j_{z\pm}$ larger than some threshold values, the energy to create pairs of spinons may vanish, leading to the condensation of spinons and a destruction of the spin liquid state. To determine this stability boundary, we need to calculate the spinon self-energy arising from the $j_{z\pm}$ term, Eq.~\eqref{eqn:int}. As in the starting point of many-body calculations, we define the non-interacting spinon Green's function:
\begin{equation}\label{eqn:g0t}
G^{(0)}(t,\boldsymbol{k})\equiv -i \langle T(\psi_{\boldsymbol{k}}(t)\psi_{\boldsymbol{k}}^\dagger(0))\rangle. 
\end{equation}
Here $T(\ldots)$ denotes the time-ordered product. Its Fourier transformation takes on the usual form \cite{mahan.2000}:
\begin{equation}\label{eqn:g0}
G^{(0)}(\omega,\boldsymbol{k})=\frac{1}{\omega^2-\omega_{\boldsymbol{k}}^2+i\delta}
\end{equation}
where $0<\delta\ll 1$. 

Even without considering the interactions with the gauge fluctuations, the spinons do interact among themselves, which we neglected in Eq.~\eqref{eqn:bosonq} as an approximation for Eq.~\eqref{eqn:hma} assuming that the boson density is low.  Our results could certainly be improved by treating these interactions as well as the exclusive nature of the bosons using standard many-body techniques. However, this is beyond the scope of the current study and will be addressed elsewhere. 

\section{Gauge fluctuations}\label{sec:gfield}

Having discussed the spinon dynamics, we proceed to consider the dynamics of the gauge field. Our description of gauge fluctuations largely follows that of Ref.~[\onlinecite{Benton.2012}]. For completeness and notational consistency, we first reproduce some of their results here. We neglect the effect of magnetic monopoles \footnote{The magnetic monopoles here are \emph{not} the same as those in classical spin ice. They arise due to the compactness of the gauge fields $A_{\boldsymbol{x}\mu}$. The interested readers are referred to Ref.~[\onlinecite{Kogut.1979}] for details.} and assume the gauge theory is in its deconfined phase. Physically, the deconfined phase corresponds to the U$(1)$ spin liquid state where it costs a finite energy to create a pair of spinons. The existence of this spin liquid state was demonstrated in Ref.~[\onlinecite{Banerjee.2008}] using quantum Monte-Carlo simulations. Under these assumptions, the Hamiltonian for the gauge sector is:
\begin{equation}\label{eqn:hg}
H_g=\sum_{\boldsymbol{x}\in\langle A\rangle,\mu}\left[\frac{U}{2}E_{\boldsymbol{x}\mu}^2+\frac{g}{2}B_{\boldsymbol{x},\mu}^2\right].
\end{equation}
The magnetic fluxes $B_{\boldsymbol{x},\mu}$ are the lattice curl of the gauge field $\tilde{A}_{\boldsymbol{x}\mu}$. $U$ and $g$ are two energy scales proportional to $j_{\pm}^3$ if only the XXZ parts of the anistropic Hamiltonian, Eqs.~\eqref{eqn:h0} and \eqref{eqn:hpm}, are considered. We assume\footnote{In Ref.~[\onlinecite{Benton.2012}], the lowest order quantum tunnelling process in quantum spin ice is translated into $2\times 12 j_{\pm}^3\cos B\approx 24 j_{\pm}^3-\frac{1}{2}\times 24 j_{\pm}^3B^2$. We identify $g_0=24 j_{\pm}^3$ here.}
\begin{equation}\label{eqn:gdef}
g=24 \zeta j_{\pm}^3\equiv g_0 \zeta 
\end{equation}
with $g_0$  as the microscopic value \cite{Hermele.2004,Benton.2012} from the third-order degenerate perturbation theory. $\zeta$ is a phenomenological factor of order $1$ which can only be determined by properly taking into account lattice scale fluctuations in a derivation starting from a microscopic model. To the best of our knowledge, such a complete microscopic construction has not yet been achieved, and is not attempted here. For simplicity, we take $\zeta=1$ from now on. 

\begin{figure}
\centering
\includegraphics[width=0.95\columnwidth]{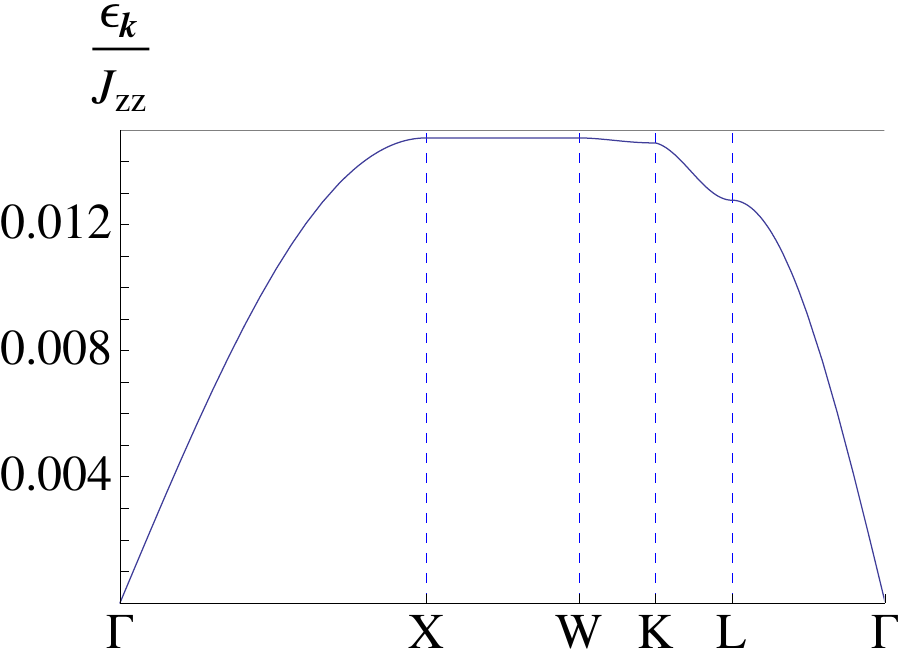}
\caption{Photon energy $\epsilon_{\boldsymbol{k}}$ along the high symmetry directions in the FCC Brillouin zone \cite{Kittel} for $j_{\pm}=0.08$. }\label{fig:lights}
\end{figure}

We follow Ref.~[\onlinecite{Benton.2012}] to quantize Eq.~\ref{eqn:hg}. We write both $B_{\boldsymbol{x}\mu}$ and $E_{\boldsymbol{x}\mu}$ in terms of Bloch modes, $B_{\boldsymbol{k}\mu}$ and $E_{\boldsymbol{k}\mu}$. $B_{\boldsymbol{k}\mu}$ is further expressed in terms of $\tilde{A}_{\boldsymbol{k}\mu}$. The magnetic energy, the second term in Eq.~\eqref{eqn:hg}, is written as the bilinear form $\tilde{A}_{\boldsymbol{k}\mu}M_{\mu\nu}\tilde{A}_{-\boldsymbol{k}\nu}$ where $M(\boldsymbol{k})$ is a Hermitian matrix,  whose explicit form is given in Appendix \ref{app:3}. We perform a unitary transformation to diagonalize $M(\boldsymbol{k})$, which results in two transverse modes $a_{j\boldsymbol{k}}$ ($j=1,2$):
\begin{equation}\label{eqn:unitrans}
\tilde{A}_{\boldsymbol{k}\mu}=\sum_{j}\eta_{\mu j}(\boldsymbol{k})a_{j\boldsymbol{k}}. 
\end{equation}
Here, $\eta_{\boldsymbol{k}}$ is a four-by-two matrix. The same unitary transformation is used to obtain the two transverse electric modes, $e_{j\boldsymbol{k}}$. After performing all these manipulations (see Ref.~[\onlinecite{Benton.2012}]), one finally obtains the Hamiltonian for the  transverse gauge fluctuations:
\begin{equation}\label{eqn:thami}
H=\sum_{\boldsymbol{k}}\sum_{j}\left[\frac{U}{2}e_{j\boldsymbol{k}}e_{j,-\boldsymbol{k}}+\frac{g}{2}\xi_{\boldsymbol{k}}^2a_{j\boldsymbol{k}}a_{j,-\boldsymbol{k}}\right]
\end{equation}
where 
\begin{equation}\label{eqn:xik}
\xi_{\boldsymbol{k}}^2=4\left(3-\frac{1}{2}\sum_{\alpha\neq\beta}\cos\frac{k_{\alpha}}{2}\cos\frac{k_{\beta}}{2}\right)\approx k^2+O(k^4). 
\end{equation}
Here $k$ is the magnitude of momentum $\boldsymbol{k}$. The Hamiltonian \eqref{eqn:xik} is a collection of non-interacting harmonic oscillators and the ``photon'' energies of the gauge fluctuations are:
\begin{equation}\label{eqn:speedoflight}
\epsilon_{\boldsymbol{k}}=\sqrt{Ug}\xi_{\boldsymbol{k}}\equiv c\xi_{\boldsymbol{k}}. 
\end{equation}
The speed of light, $c$, has been measured using quantum Monte-Carlo simulations \cite{Benton.2012} to be $c\sim 0.3 g_0$. The photon dispersion $\xi_{\boldsymbol{k}}$ is illustrated in Fig.~\ref{fig:lights}. Using Eqs.~\eqref{eqn:gdef} and \eqref{eqn:speedoflight}, we can extract the value of $U$ to be 
\begin{equation}\label{eq:udef}
U=\frac{0.09 g_0}{\zeta}=\frac{2.16 j_{\pm}^3}{\zeta}. 
\end{equation}
We observe that photon energy $\epsilon_{\boldsymbol{k}}$ is proportional to the spinon energy $\omega_{\boldsymbol{k}}$ at the condensation point of spinons, $j_{\pm}=1/12$.

As in Section \ref{sec:spinon}, we define the following Green's function \cite{mahan.2000} for the transverse electric fluxes:
\begin{equation}\label{eqn:efluxt}
g_{ij}(t,\boldsymbol{k})=-i\langle T(e_{i\boldsymbol{k}}(t)e_{j,-\boldsymbol{k}}(0))\rangle\equiv \delta_{ij}g(t,\boldsymbol{k}). 
\end{equation}
The Fourier transform of $g(t,\boldsymbol{k})$ is:
\begin{equation}\label{eqn:eflux}
g(\omega,\boldsymbol{k})=\frac{\epsilon_{\boldsymbol{k}}^2}{U(\omega^2-\epsilon_{\boldsymbol{k}}^2+i\delta)}. 
\end{equation}

Having discussed the dynamics of both spinon fields and gauge fields separately, we now move on to consider their interaction. 

\section{The spinon self-energy and the phase diagram}\label{sec:senergy}
Our main goal now is to estimate the stability of the U$(1)$ liquid phase with a finite $j_{z\pm}$. In Section \ref{sec:spinon}, we showed that the spinon energy $\omega_{\boldsymbol{k}}$ vanishes at $\boldsymbol{k}=0$ when $j_{\pm}=1/12$. The U$(1)$ liquid then becomes unstable toward the formation of an XY antiferromagnetic phase. Since $j_{z\pm}$ couples electric field flux $E_{\boldsymbol{x}\mu}$ with two spinon fields, it changes the spinon dispersion through a self-energy contribution. For $j_{z\pm}$ larger than a threshold value, which defines a stability boundary, the renormalized spinon energy vanishes, leading to their condensation. To expose this stability phase diagram of the Hamiltonian \eqref{eqn:hami} in the $j_{\pm}$ vs $j_{z\pm}$ plane, we calculate the lowest order self-energy contribution from the interaction between spinons and transverse gauge fluctuations. Neglecting all higher order terms involving $\tilde{A}_{\boldsymbol{x}\mu}$, the lowest order coupling from Eq.~\eqref{eqn:int} reads:
\begin{equation}\label{eqn:coupling}
\begin{aligned}
H_{\mathrm{int}}\approx &-\frac{j_{z\pm}}{2}\sum_{\boldsymbol{x}}\sum_{\mu\neq\nu}\left[E_{\boldsymbol{x}\mu}(\psi_{\boldsymbol{x}}^\dagger\psi_{\boldsymbol{x}+\hat{\nu}}+\psi^\dagger_{\boldsymbol{x}+\hat{\mu}-\hat{\nu}}\psi_{\boldsymbol{x}+\hat{\mu}})\mathrm{e}^{i\phi_{\mu\nu}}\right.\\
&\left.+h.c\right].
\end{aligned}
\end{equation}
Using the standard operator formulation of many-body theory \cite{mahan.2000}, the lowest order correction to the spinon Green's function is:
\begin{widetext}
\begin{equation}\label{eqn:g1}
\begin{aligned}
G^{(1)}(t,\boldsymbol{k})=i\left\langle T\left(\psi_{\boldsymbol{k}}(t)\psi_{\boldsymbol{k}}^\dagger(0)\int_{-\infty}^{\infty}dt_1dt_2\,H_{\mathrm{int}}(t_1)H_{\mathrm{int}}(t_2)\right)\right\rangle 
\end{aligned}
\end{equation}
\end{widetext}
where all operators are written in the interaction picture. The simplest scheme to take into account the coupled dynamics of two interacting quantum fields is the random phase approximation (RPA) \cite{mahan.2000}. Here we use the RPA approximation to describe the effect of the gauge field on the spinons. Under RPA, the full Green's function is:
\begin{equation}
[G(\omega,\boldsymbol{k})]^{-1}=[G^{0}(\omega,\boldsymbol{k})]^{-1}-\Sigma(\omega,\boldsymbol{k}). 
\end{equation}
To compute the self-energy $\Sigma(\omega,\boldsymbol{k})$, we use Wick's theorem \cite{mahan.2000} to write Eq.~\eqref{eqn:g1} as a convolution of non-interacting spinon Green's function $G^{(0)}(t,\boldsymbol{k})$, Eq.~\eqref{eqn:g0t} and electric flux Green's function $g(t,\boldsymbol{k})$, Eq.~\eqref{eqn:efluxt}. The convolution becomes a product as we transform all Green's functions into the frequency space.  After these standard procedures \cite{mahan.2000}, the self-energy contribution, $\Sigma(\omega,\boldsymbol{k})$, proportional to $j_{z\pm}^2$, is:
\begin{equation}
\begin{aligned}
\Sigma(\omega,\boldsymbol{k})=&\frac{ij_{z\pm}^2}{2N}\sum_{\mu,\nu,j}\sum_{\boldsymbol{k}}\eta_{\mu j}(\boldsymbol{q})\eta_{\nu j}(\boldsymbol{q})f_{\mu\nu}(\boldsymbol{k}-\boldsymbol{q},\boldsymbol{k})\\
&\int \frac{d\omega^\prime}{2\pi}g(\omega^\prime,\boldsymbol{q})G^{(0)}(\omega-\omega^\prime,\boldsymbol{k}-\boldsymbol{q}). 
\end{aligned}
\end{equation}
We note that only the lowest order approximation in $j_{z\pm}$ of the self energy is included here. The approximation is only valid for small $j_{z\pm}$. The explicit form of $f(\boldsymbol{k}_1,\boldsymbol{k}_2)$ is given in Appendix \ref{app:3}.  Expanding $\Sigma(\omega,\boldsymbol{k})$ around $\omega=\omega_{\boldsymbol{k}}$, the energy of coherent spinons, $\omega_{\boldsymbol{k}}$, is corrected by the $\Sigma(\omega,\boldsymbol{k})$, approximated by $\Sigma(\omega_{\boldsymbol{k}},\boldsymbol{k})$, to order $j_{z\pm}^2$:
\begin{equation}\label{eqn:newfreq}
\tilde{\omega}_{\boldsymbol{k}}^2\approx \omega_{\boldsymbol{k}}^2+\Sigma(\omega_{\boldsymbol{k}},\boldsymbol{k}).
\end{equation}
All corrections to Eq.~\eqref{eqn:newfreq} is of fourth power of $j_{z\pm}$ or higher. Since all known long-range ordered phases beyond the stability region of the U$(1)$ spin liquid, including the splayed ferromagnetic (SFM) phase \cite{Yaouanc.2013,Wong.2013} and the antiferromagnetic XY order \cite{Champion.2003,Wong.2013}, are translationally invariant, we focus on the spinon energy $\tilde{\omega}(\boldsymbol{k})$ at $\boldsymbol{k}=0$. 

\begin{figure}
\centering
\includegraphics[width=0.9\columnwidth]{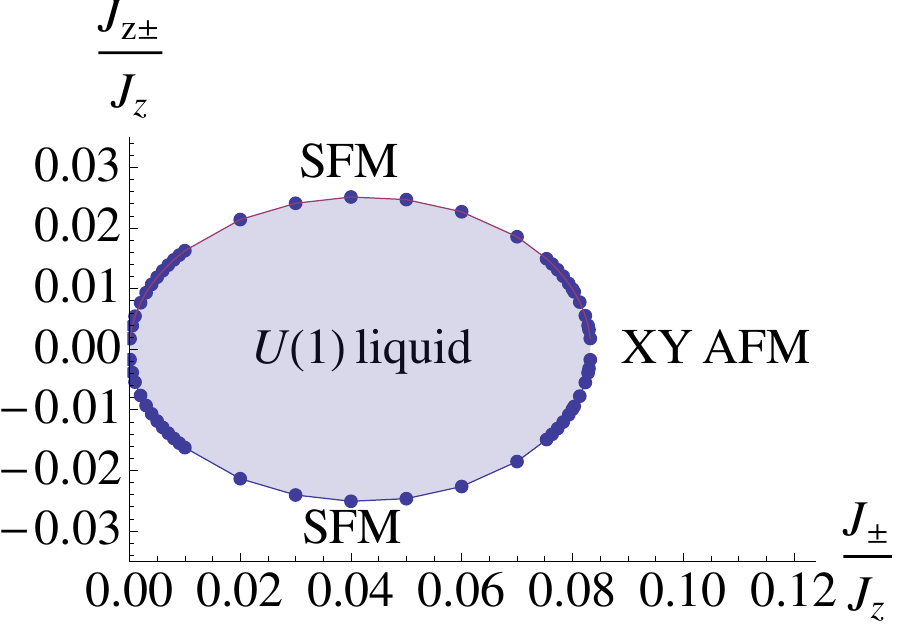}\\
\includegraphics[width=0.9\columnwidth]{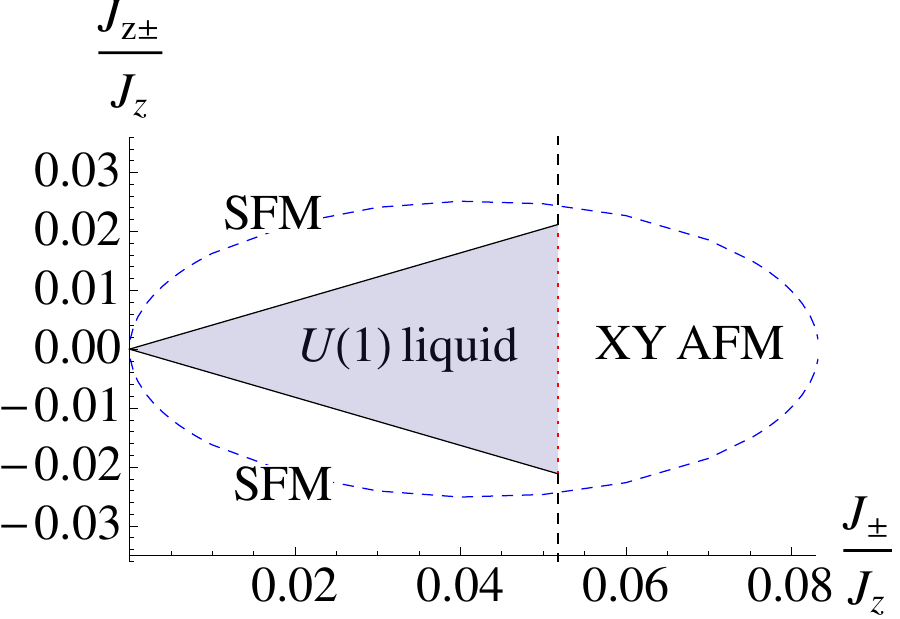}
\caption{The stability region of the U(1) liquid phase (top) and the estimated phase diagram (bottom). The phase boundary between the U$(1)$ liquid phase and the XY antiferromagnetic phase (vertical red dotted line), labelled as XY AFM, is determined based on the results of quantum Monte-Carlo simulation in Ref.~[\onlinecite{Banerjee.2008}]. The blue dashed oval in the bottom panel is the perturbative stability region (shown in the top panel) of the U(1) liquid phase. The phase boundaries between the splayed ferromagnetic \cite{Yaouanc.2013} (SFM) phase and the XY antiferromagnetic phase have not been determined in this work. Consequently, we do not draw a boundary between these phases in the top panel. See Ref.~[\onlinecite{Savary.2012,Wong.2013,Yan.2013}] for estimations of the phase boundaries between the SFM phase and the XY antiferromagnet long-range ordered phases.}\label{fig:pd}
\end{figure}

We calculate $\tilde{\omega}_{\boldsymbol{k}=0}^2$ for different $j_{\pm}$ and $j_{z\pm}$ by computing $\Sigma(\omega_{\boldsymbol{k}},\boldsymbol{k})$ numerically. For sufficiently small $j_{z\pm}$, the energy of the zero momentum spinon is reduced but remains finite. At a threshold value $j_{z\pm}=j_{c}(j_{\pm})$, $\tilde{\omega}_{\boldsymbol{k}=0}$ become zero and the $\boldsymbol{k}=0$ spinon condenses. Since the correction to self-energies only comes in as even powers of $j_{z\pm}$, the stability boundary of the U$(1)$ liquid is symmetric under $j_{z\pm}\to -j_{z\pm}$ (Fig.~\ref{fig:pd}). For $|j_{z\pm}|>j_{c}$, the system is expected to order in one of the adjacent long-range ordered phases, either the SFM phase or the antiferromagnetic XY order (Fig.~\ref{fig:pd}). Note that we did not determine the classical phase boundary between the SFM phase and the antiferromagnetic XY order here. The reader can refer to Refs.~[\onlinecite{Savary.2012,Wong.2013,Yan.2013}] for a determination, within a classical approximation, of the phase boundaries between these conventional long-range ordered phases.

\begin{figure}
\includegraphics[width=0.95\columnwidth]{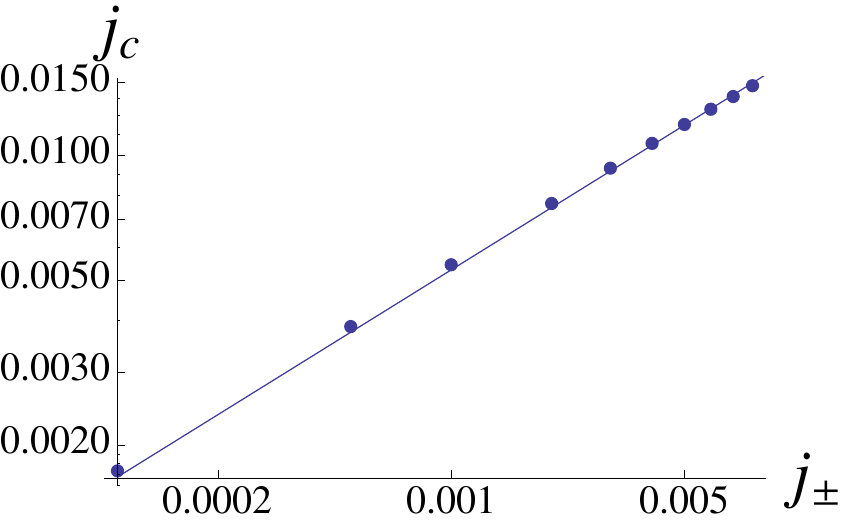}\\
\includegraphics[width=0.95\columnwidth]{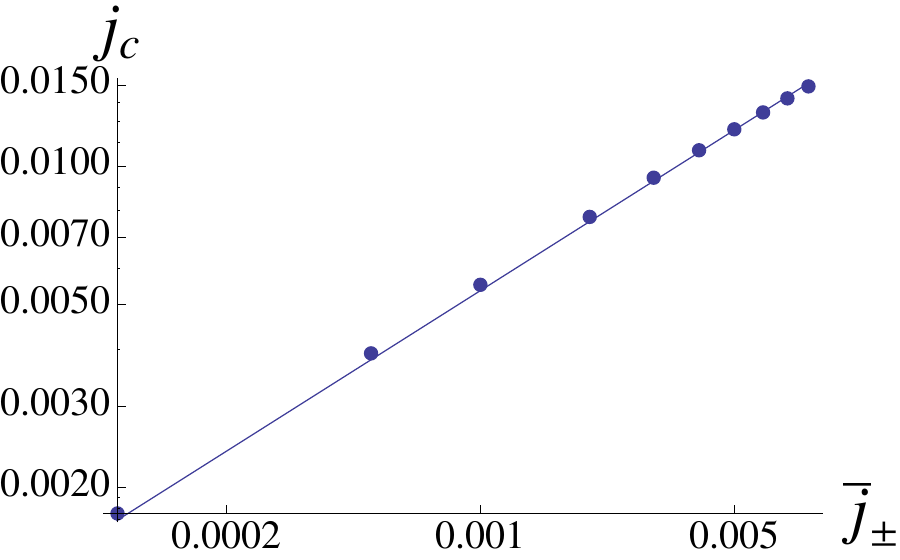},
\caption{The limiting behavior of the stability boundary $j_{c}(j_{\pm})$ for $j_{\pm}\to 0$ (top) and $j_{\pm}\to 1/12$ (bottom). $\bar{j}_\pm\equiv 1/12-j_{\pm}$. The numerical datas are plotted against fitted forms, $0.167 j_{\pm}^{1/2}$ and $0.169 \bar{j}_\pm^{1/2}$, respectively. }\label{fig:check}
\end{figure}

Let us comment on the asymptotic behavior of the stability boundary for small $j_{\pm}$ and for $j_{\pm}$ close to $1/12$, the critical value for spinon to condense for $j_{z\pm}=0$, found in Section \ref{sec:spinon}. For $j_{\pm}\ll 1$, the correction to the $\boldsymbol{k}=0$ spinon gap is expected to scale as $j_{z\pm}^2/j_{\pm}$ from simple second-order Rayleigh-Schr\"{o}dinger perturbation theory. Since the gap in the limit of $j_{\pm}\to 0$ is of the order $J_{zz}$, the stability boundary is thus expected to be of form $j_{z\pm}\sim \pm\sqrt{j_{\pm}}$ as $j_{\pm}\to 0$. On the other hand, $\omega_{\boldsymbol{k}=0}^2$ vanishes as $|1/12-j_{\pm}|$ while $\Sigma(\omega_{\boldsymbol{k}},\boldsymbol{k})$ approaches $c_0 j_{z\pm}^2$, $c_0$ being a constant, as $j_{\pm}\to 1/12$.  We thus expect the stability boundary to behave as $j_{z\pm}\sim \pm \sqrt{1/12-j_{\pm}}$ for $j_{\pm}$ close to $1/12$. Both expectations are explicitly verified in our numerical results for $j_c(j_{\pm})$ (Fig. \ref{fig:check}). 

The above results rely on a stability analysis to determine the phase boundary between the U$(1)$ liquid phase and semi-classically ordered phases. In practice, such phase transitions may be preempted by a first-order transition\cite{Makhfudz.2014}. In fact, a quantum Monte-Carlo study \cite{Banerjee.2008, Kato.2014} of the XXZ model on the pyrochlore lattice finds that the U$(1)$ liquid phase undergoes a first-order phase transition \cite{Banerjee.2008} at $j_{\pm}\approx 0.05$, smaller than our estimate of spinon condensation threshold $j_{\pm}=1/12\approx 0.083$. The phase boundary between the U$(1)$ liquid phase and the SFM phase can be roughly estimated by comparing the energies of the two phases to lowest order in $j_{z\pm}$ and $j_{\pm}$. To lowest order in $j_{z\pm}$, the energy per FCC unit cell for the SFM phase is:
\begin{equation}
E_{\mathrm{sfm}}=-4\times 3\times 3 S^2j_{z\pm}^2=-9j_{z\pm}^2
\end{equation}
where $S=1/2$ is the spin length. Here the factor $4$ counts the number of sites in the unit cell on the pyrochlore lattice, one factor of $3$ is the number of third nearest-neighbor of the ``$a$-type'' for a site, $6$, divided by $2$. We note that there are two types of third nearest neighbors on the pyrochlore lattice \cite{Gardner_RMP}. The bond connecting a site and its third nearest neighbor of ``$a$-type'' goes through another pyrochlore site. Finally, $-3j_{z\pm}^2$ is the lowest order perturbative contribution \cite{Ross.2011,Applegate.2012} to exchange interaction between a site and its third nearest-neighbor of ``$a$-type''. 
For the U$(1)$ liquid, the lowest order energy in $j_{\pm}$ per unit cell comes from the zero point energy in the spinon sector (Eq.~\eqref{eqn:e0}):
\begin{equation*}
E_0\approx -\frac{3j_{\pm}^2}{2}. 
\end{equation*}
Using these estimates, the phase boundary between the U$(1)$ liquid and the SF phase is determined by $j_{z\pm}\approx\pm j_{\pm}/\sqrt{6}$. Both the stability region of the U(1) liquid phase and the estimated phase diagram determined above are displayed in Fig.~\ref{fig:pd}

\section{Discussion}\label{sec:dis}
In this work, we developed a formalism to study quantum spin liquids in highly anisotropic spin models on the pyrochlore lattice. Building on previous works, we considered the dynamics in both spinons and gauge sectors. By applying our formalism to a particular anisotropic model, Eq.~\eqref{eqn:hami}, we mapped out the stability region of the U$(1)$ liquid phase in the $j_{\pm}$ vs $j_{z\pm}$ plane and estimated the phase boundaries between the spin liquid phase and close-by conventional long-range magnetically ordered phases. Our formalism can be readily applied to the study of other possible spin liquid phases as long as they are the descendants of the classical spin ice solution. 

It is interesting to discuss the general merit of the formalism we developed in this work. As stated in Section \ref{sec:gfield}, our description of the transverse gauge fluctuations is not ideal since it is not derived from a full microscopic theory. However, it was demonstrated, using large scale quantum Monte-Carlo simulation \cite{Benton.2012}, that the theory describes the simulation data quantitatively. While it is well known \cite{Hermele.2004,Benton.2012,Castro.2006} that $g$ in the Hamiltonian \eqref{eqn:hg} is of the order $j_{\pm}^3$ for the XXZ model, it is perhaps a bit surprising that $U$ acquires almost full scale of $24j_{\pm}^3$ (Eq.~\eqref{eq:udef}). In the conventional presentation of the theory \cite{Hermele.2004,Benton.2012,Castro.2006}, $U$ is supposed to be very large such that $E_{\boldsymbol{x}\mu}=\pm 1/2$, consistent with the spin-length, is enforced. That $U\sim 24j_{\pm}^3$ can be understood physically by accepting that the U$(1)$ compact gauge theory is an inexact mapping of the spin model. Within the spin ice manifold, the lowest order quantum tunnelling operator, which flips six alternating spins around a hexagonal plaquette, does not commute if two hexagonal plaquettes share spins. On the contrary, such operators translate into $\cos(B)$ to lowest order on each plaquette within the gauge theory formulation, which commute with each other. The $UE^2/2$ term has to be present in order to preserve the quantum nature of the theory. From this perspective, since the amplitude $g\sim j_{\pm}^3$ of the quantum tunnelling operator is the only energy scale within the classical spin-ice manifold, it is perhaps natural to expect that $U$ is dynamically generated such that $U\sim g$. 

As the gauge field, and its associated emergent photon is one of the key feature of the $U(1)$ theory description of the quantum spin ice state, it is useful to ascribe a physical meaning to the gauge fluctuations. The photon, as in standard quantum electrodynamics, contains both electric and magnetic parts. In quantum spin ice, electric fluxes are the Ising components of the spins. This implies that the electric part of the emerging photon are fluctuations of the local Ising spin components. The collective excitation has a vector amplitude whose direction is perpendicular to its propagating direction defined by its momentum $\boldsymbol{q}$. In other words, it is ``transverse'' just like real photons. The interpretation is supported by the connection of quantum spin ice to a quantum dimer model where emergent photons are known to be fluctuations of dimer densities \cite{ LMMqdmchapter,Huse.2003,Moessner.2003}. The hardcore dimers correspond to, for example, $S_{\boldsymbol{x}\mu}^{z}=1/2$ in spin ice. Transverse dimer density fluctuations are thus transverse fluctuations of local Ising components of the spins, or transverse magnetization waves.  

We would like to comment on the similarity and differences between the large-$N$ approach of Ref.~[\onlinecite{Savary.2012}] and our approximation based on the exclusive boson representation of an XY quantum rotor. In both formalism, the spinons are represented by bosonic degrees of freedom moving under a static background gauge field. The ground state wave functions obtained in both approximation contains basis states that violate the Gauss law \eqref{eqn:charged}. Proper projectors would need to be applied to address this issue. There are also a couple of differences. In the large-$N$ approximation, the constrain $|\psi_{\boldsymbol{x}}|=1$ is enforced only on average. The real and imaginary parts of $\psi_{\boldsymbol{x}}$ can be interpreted as the coordinates $(x,y)$ of a two-dimensional particle. The large-$N$ approximation \cite{Savary.2012} frees the particle to move on the entire two-dimensional plane instead of the unit circle $\sqrt{x^2+y^2}=1$. As a result, the momentum of the particle, which translates to the charge $\hat{Q}$ (Ref.~[\onlinecite{Savary.2012}]), can take on continuous value. In our approximation, $\hat{Q}=n_d-n_b$ takes only discrete values. Most of the quantitative differences about spinon dynamics between our work and Ref.~[\onlinecite{Savary.2012}] stems from this distinction. Furthermore, in the large-$N$ approximation, the Lagrange multiplier $\lambda$ needed to enforce the global constrain $\sum_{\boldsymbol{x}}[|\psi_{\boldsymbol{x}}|^2-1]$ needs to be computed self-consistently for a given set of exchange parameters. No such computations are needed in the lowest order approximation of the exclusive boson formulation. 

While we applied the exclusive boson formalism to the quantum spin ice problem, we stress that it is a general representation of the XY quantum rotor. Starting from the formalism, standard diagrammatic techniques could be applied to XY quantum rotor models. As the boson density $n$ increase, interactions among bosons as well as their exclusiveness become important. The naive approximation where only operators quadratic in bosons are kept is bound to fail under such circumstances. However, even in such cases, the applicability of the formalism could perhaps be greatly enhanced if used in combination with sophisticated numerical methods such as variational quantum Monte Carlo. We also note that the XY quantum rotors are used to represent the charge degrees of freedom in weak Mott insulators \cite{Florens.2004}. Perhaps the exclusive boson formalism could prove useful in such a context as well. 

In the future, one could apply the present formalism to study the zero-temperature phase diagram of the frustrated XXZ model on the pyrochlore lattice, i.e. Eq.~ \eqref{eqn:h0} and \eqref{eqn:hpm} in the case where $J_{\pm}<0$. It would also be interesting to study the phase diagram of systems with both Kramers and non-Kramers magnetic ions and taking into account a finite $J_{\pm\pm}$ coupling. Moreover, one could extend the present formalism to consider finite temperature properties of quantum spin ice with the aim of exploring the physics of quantum spin ice candidates such as Yb$_2$Ti$_2$O$_7$, Tb$_2$Ti$_2$O$_7$, Pr$_2$Zr$_2$O$_7$ and Pr$_2$Sn$_2$O$_7$.

\begin{acknowledgments}
We thank Leon Balents, Lucile Savary and Oleg Tchernyshyov for useful discussions and comments on an earlier version of the manuscript. This research was supported by the NSERC of Canada, the Canada Research Chair program (M.G., Tier 1). Research at the Perimeter Institute for Theoretical Physics is supported by the Government of Canada through Industry Canada and by the Province of Ontario through the Ministry of Economic Development and Innovation.
\end{acknowledgments}

\begin{appendix}
\section{Local coordinate systems}\label{app:1}
For completeness, we list the local coordinates for the four sublattices of the pyrochlore lattice. We use $\hat{x}_{\alpha}$, $\hat{y}_{\alpha}$ and $\hat{z}_{\alpha}$ for the local axes with $\alpha=0,1,2,3$. The global cartesian axes are labeled as $\hat{x}$, $\hat{y}$ and $\hat{z}$. We follow the convention of Ref.~[\onlinecite{Ross.2011}].
\begin{subequations}
\begin{eqnarray}
&&\left\{\begin{array}{l}\hat{x}_0=\frac{1}{\sqrt{6}}(-2\hat{x}+\hat{y}+\hat{z})\\ \hat{y_0}=\frac{1}{\sqrt{2}}(-\hat{y}+\hat{z})\\ \hat{z_0}=\frac{1}{\sqrt{3}}(\hat{x}+\hat{y}+\hat{z})\end{array}\right., \\
&&\left\{\begin{array}{l}\hat{x}_1=\frac{1}{\sqrt{6}}(-2\hat{x}-\hat{y}-\hat{z})\\ \hat{y_1}=\frac{1}{\sqrt{2}}(\hat{y}-\hat{z}) \\ \hat{z_1}=\frac{1}{\sqrt{3}}(\hat{x}-\hat{y}-\hat{z})\end{array}\right., \\
&&\left\{\begin{array}{l}\hat{x}_2=\frac{1}{\sqrt{6}}(2\hat{x}+\hat{y}-\hat{z})\\ \hat{y_2}=\frac{1}{\sqrt{2}}(-\hat{y}-\hat{z})\\ \hat{z_2}=\frac{1}{\sqrt{3}}(-\hat{x}+\hat{y}-\hat{z})\end{array}\right., \\
&&\left\{\begin{array}{l}\hat{x}_3=\frac{1}{\sqrt{6}}(2\hat{x}-\hat{y}+\hat{z})\\ \hat{y_3}=\frac{1}{\sqrt{2}}(\hat{y}+\hat{z})\\ \hat{z_3}=\frac{1}{\sqrt{3}}(-\hat{x}-\hat{y}+\hat{z})\end{array}\right. . 
\end{eqnarray}
\end{subequations}

We also give explicitly $\hat{\mu}$'s:
\begin{subequations}
\begin{eqnarray}
\hat{0}&=&\frac{1}{4}(\hat{x}+\hat{y}+\hat{z}),\\
\hat{1}&=&\frac{1}{4}(\hat{x}-\hat{y}-\hat{z}),\\
\hat{2}&=&\frac{1}{4}(\hat{y}-\hat{x}-\hat{z}),\\
\hat{3}&=&\frac{1}{4}(\hat{z}-\hat{x}-\hat{y}). 
\end{eqnarray}
\end{subequations}

\section{The Bogoliubov transformation}\label{app:bogo}
In this appendix, we give the explicit form of the Bogoliubov transformation that leads to Eq.~\eqref{eqn:dispersion}. In terms of Bloch modes $d_{\boldsymbol{k}}$ and $b_{\boldsymbol{k}}$, Eq.~\eqref{eqn:bosonq} can be written as:
\begin{widetext}
\begin{equation}\label{eqn:ee1}
H=\sum_{\boldsymbol{k}}\left[\left(\frac{1}{2}-\frac{j_{\pm}}{2}\sum_{\alpha\neq\beta}\cos\frac{k_{\alpha}}{2}\cos\frac{k_{\beta}}{2}\right)(d_{\boldsymbol{k}}^\dagger d_{\boldsymbol{k}}+b_{-\boldsymbol{k}} d_{-\boldsymbol{k}}^\dagger)-\frac{j_{\pm}}{2}\sum_{\alpha\neq\beta}\cos\frac{k_{\alpha}}{2}\cos\frac{k_{\beta}}{2}(d_{\boldsymbol{k}}^\dagger b_{-\boldsymbol{k}}^\dagger+h.c)-\frac{1}{2}\right]
\end{equation}
\end{widetext}
To diagonalize the Hamiltonian, we perform a Bogoliubov transformation:
\begin{subequations}
\begin{eqnarray}
d_{\boldsymbol{k}}&=&\tilde{d}_{\boldsymbol{k}}\cosh\gamma_{\boldsymbol{k}}+\tilde{b}_{-\boldsymbol{k}}^\dagger \sinh\gamma_{\boldsymbol{k}},\\
b_{\boldsymbol{k}}&=&\tilde{b}_{\boldsymbol{k}}\cosh\gamma_{\boldsymbol{k}}+\tilde{d}_{-\boldsymbol{k}}^\dagger \sinh\gamma_{\boldsymbol{k}}.
\end{eqnarray}
\end{subequations}
$\gamma_{\boldsymbol{k}}$ satisfies:
\begin{subequations}
\begin{eqnarray}
\cosh2\gamma_{\boldsymbol{k}}&=&\frac{1}{\omega_{\boldsymbol{k}}}\left(\frac{1}{2}-\frac{j_{\pm}}{2}\sum_{\alpha\neq\beta}\cos\frac{k_{\alpha}}{2}\cos\frac{k_{\beta}}{2}\right) \\
\sinh2\gamma_{\boldsymbol{k}}&=&\frac{j_{\pm}}{2\omega_{\boldsymbol{k}}}\sum_{\alpha\neq\beta}\cos\frac{k_{\alpha}}{2}\cos\frac{k_{\beta}}{2}
\end{eqnarray}
\end{subequations}
with $\omega_{\boldsymbol{k}}$ given in Eq.~\eqref{eqn:dispersion}. In terms of quasi-particles $\tilde{b}$, and $\tilde{d}$, Eq.~\eqref{eqn:ee1} reads:
\begin{equation}
H=\sum_{\boldsymbol{k}}\left[\omega_{\boldsymbol{k}}(\tilde{d}_{\boldsymbol{k}}^\dagger \tilde{d}_{\boldsymbol{k}}+\tilde{b}_{\boldsymbol{k}}^\dagger \tilde{b}_{\boldsymbol{k}}+1)-\frac{1}{2}\right]. 
\end{equation}

The average number of spinons per site, $n$, can be computed as:
\begin{equation}\label{eq:ndensity}
n=\frac{1}{V}\sum_{\boldsymbol{k}}\langle b_{\boldsymbol{k}}^\dagger b_{\boldsymbol{k}}+d_{\boldsymbol{k}}^\dagger d_{\boldsymbol{k}}\rangle=\frac{2}{V}\sum_{\boldsymbol{k}}\left[\frac{1}{2}\cosh 2\gamma_{\boldsymbol{k}}-\frac{1}{2}\right] 
\end{equation}
where $V$ is the number of FCC unit cells. Using Eq.~\eqref{eq:ndensity}, $n$ can be computed numerically for any given $j_{\pm}$. 

\section{The explicit form for $M(\boldsymbol{k})$ and $f_{\boldsymbol{k}_1,\boldsymbol{k}_2}$}\label{app:3}

Here, we give the explicit form for $M(\boldsymbol{k})$ and $f_{\boldsymbol{k}_1,\boldsymbol{k}_2}$ used in Section \ref{sec:gfield} and \ref{sec:senergy}, respectively. $M$ is a Hermitian matrix so that we list only $M_{\mu\nu}(\boldsymbol{k})$ with $\mu\ge \nu$. 
\begin{subequations}
\begin{eqnarray}
M_{00}(\boldsymbol{k})&=&6-2\left[\cos\frac{k_x-k_y}{2}+\cos\frac{k_x-k_z}{2}+\cos\frac{k_y-k_z}{2}\right],\nonumber\\
M_{11}(\boldsymbol{k})&=&6-2\left[\cos\frac{k_x+k_y}{2}+\cos\frac{k_y-k_z}{2}+\cos\frac{k_x+k_z}{2}\right],\nonumber\\
M_{22}(\boldsymbol{k})&=&6-2\left[\cos\frac{k_x+k_y}{2}+\cos\frac{k_x-k_z}{2}+\cos\frac{k_y+k_z}{2}\right],\nonumber\\
M_{33}(\boldsymbol{k})&=&6-2\left[\cos\frac{k_x-k_y}{2}+\cos\frac{k_y+k_z}{2}+\cos\frac{k_x+k_z}{2}\right],\nonumber\\
M_{01}(\boldsymbol{k})&=&2\left[\mathrm{e}^{\frac{ik_y}{2}}\cos\frac{k_{x}}{2}+\mathrm{e}^{\frac{ik_{z}}{2}}\cos\frac{k_x}{2}-1-\mathrm{e}^{\frac{i(k_y+k_z)}{2}}\right],\nonumber\\
M_{02}(\boldsymbol{k})&=&2\left[\mathrm{e}^{\frac{ik_x}{2}}\cos\frac{k_{y}}{2}+\mathrm{e}^{\frac{ik_{z}}{2}}\cos\frac{k_y}{2}-1-\mathrm{e}^{\frac{i(k_x+k_z)}{2}}\right],\nonumber\\
M_{03}(\boldsymbol{k})&=&2\left[\mathrm{e}^{\frac{ik_x}{2}}\cos\frac{k_{z}}{2}+\mathrm{e}^{\frac{ik_{y}}{2}}\cos\frac{k_z}{2}-1-\mathrm{e}^{\frac{i(k_x+k_y)}{2}}\right],\nonumber\\
M_{12}(\boldsymbol{k})&=&2\left[\mathrm{e}^{\frac{ik_x}{2}}\cos\frac{k_{z}}{2}+\mathrm{e}^{\frac{-ik_{y}}{2}}\cos\frac{k_z}{2}-1-\mathrm{e}^{\frac{i(k_x-k_y)}{2}}\right],\nonumber\\
M_{13}(\boldsymbol{k})&=&2\left[\mathrm{e}^{\frac{ik_x}{2}}\cos\frac{k_{y}}{2}+\mathrm{e}^{\frac{-ik_{z}}{2}}\cos\frac{k_y}{2}-1-\mathrm{e}^{\frac{i(k_x-k_z)}{2}}\right],\nonumber\\
M_{23}(\boldsymbol{k})&=&2\left[\mathrm{e}^{\frac{ik_y}{2}}\cos\frac{k_{x}}{2}+\mathrm{e}^{\frac{-ik_{z}}{2}}\cos\frac{k_x}{2}-1-\mathrm{e}^{\frac{i(k_y-k_z)}{2}}\right].\nonumber
\end{eqnarray}
\end{subequations}

$f(\boldsymbol{k}_1,\boldsymbol{k}_2)$ is also a $4\times 4$ matrix. Its matrix elements are:
\begin{equation*}
\begin{aligned}
&f_{\mu\nu}(\boldsymbol{k}_1,\boldsymbol{k}_2)=\sum_{\mu^\prime}^{\mu^\prime\neq\mu}\sum_{\nu^\prime}^{\nu^\prime\neq \nu}\mathrm{e}^{i(\phi_{\mu\mu^\prime}-\phi_{\nu\nu^\prime})}\left[\mathrm{e}^{-i\boldsymbol{k}_1\cdot\hat{\mu}^\prime}+\right.\\
&\left.\mathrm{e}^{i[\boldsymbol{k}_2\cdot(\hat{\mu}-\hat{\mu}^\prime)-\boldsymbol{k}_1\cdot\hat{\mu}]}\right]\left[\mathrm{e}^{i\boldsymbol{k}_1\cdot\hat{\nu}^\prime}+\mathrm{e}^{-i[\boldsymbol{k}_2\cdot(\hat{\nu}-\hat{\nu}^\prime)-\boldsymbol{k}_1\cdot\hat{\nu}]}\right].
\end{aligned}
\end{equation*}

\section{The dispersion of a single spinon: a variational estimation}\label{app:4}
Projecting onto the states with two spinons of opposite charges, Eq.~\eqref{eqn:hma} describes only the hopping dynamics of a spinon if the other spinon's position is fixed and the scattering between the two spinons is neglected. Without losing any generality, we assume the spinon with negative charge is fixed while the positive spinon on sublattice A is mobile. By applying $H$ on a state $|\boldsymbol{x}\rangle$ with a monopole on site $\boldsymbol{x}$, the equation of motion for the positive spinon is: 
\begin{equation}
H|\boldsymbol{x}\rangle=\frac{1}{2}|\boldsymbol{x}\rangle-\frac{j_{\pm}}{4}\sum_{\mu\neq\nu}|\boldsymbol{x}+\hat{\mu}-\hat{\nu}\rangle. 
\end{equation}

The normal modes of the mobile spinon can be obtained by a simple Fourier transformation:
\begin{equation}
H|\boldsymbol{k}\rangle=\left[\frac{1}{2}-\frac{j_{\pm}}{2}\sum_{\alpha\neq\beta}\cos\frac{k_{\alpha}}{2}\cos\frac{k_{\beta}}{2}\right]|\boldsymbol{k}\rangle. 
\end{equation}
The energy agrees  with Eq.~\eqref{eqn:app}. It does not agree with Eq.~\eqref{eqn:dispersion} since we neglect the creation and annihilation processes of spinons which are captured by the Bogoliubov transformation leading to Eq.~\eqref{eqn:dispersion}. 

\section{The CFM phase within the g-MFT formalism}\label{app:5}
We expect the XY components of the spins to have a static expectation value in the CFM phase. To first see this, we step back from the gauge theory picture and consider the original spin model. Consider a state with ordered Ising moments  $S_{\boldsymbol{x}\mu}^z$ with static expectation value $m_z\sim \langle S_{\boldsymbol{x}\mu}^z\rangle$ but with fluctuating transverse (XY) components. The (free) energy of the system can be expanded as a function of small $m_{\perp}$:
\begin{equation*}
F=r m_z m_{\perp}+b m_{\perp}^2+\ldots
\end{equation*}
Here $b$ is positive as we assume that the transverse moments do not spontaneously order. We stress that both $m_z$ and $m_{\perp}$ should be understood as linear combination of $\langle S_{\boldsymbol{x}\mu}^z\rangle$ and $|\langle S_{\boldsymbol{x}\mu}^+\rangle|$ with proper symmetry properties. The qualitative argument is clear, we believe, without constructing the exact Landau functional. The linear coupling between $m_{z}$ and $m_{\perp}$, $r$, is proportional to $j_{z\pm}$. Clearly, for $m_z\neq 0$, $F$ is minimized for a nonzero $m_{\perp}$, inevitably. 

We now demonstrate that the CFM phase, with properties such as defined in Ref.~[\onlinecite{Savary.2012}], has both ordered Ising \emph{and} transverse components even \emph{within} the gauge mean-field formalism \cite{Savary.2012}. We recall that in the gauge mean-field theory, the expectation value of the transverse component $S^{+}_{\boldsymbol{x}\mu}$, is:
\begin{equation}\label{eqn:sp}
\langle S^{+}_{\boldsymbol{x}\mu}\rangle=\langle s_{\boldsymbol{x}\mu}^+\rangle\langle \psi_{\boldsymbol{x}}^\dagger\psi_{\boldsymbol{x}+\hat{\mu}}\rangle. 
\end{equation}
The second bracket of \eqref{eqn:sp} is the average of inter-sublattice correlation of spinons. Per its self-consistent mean-field solution, the CFM phase is defined by the following properties \cite{Savary.2012}:
\begin{equation}\label{eqn:a1}
\langle s^+_{\boldsymbol{x}\mu}\rangle\neq0, \qquad \langle S^{z}_{\boldsymbol{x}\mu}\rangle\neq 0. 
\end{equation}
The product of the two expectation values leads to nonzero amplitude for spinons to tunnel from one sublattice to the other, as is implicit in Eq.~\ref{eqn:hb}. In general, the system takes advantage of the tunneling process by developing finite intersublattice correlations:
\begin{equation}\label{eqn:a2}
\langle \psi_{\boldsymbol{x}}^\dagger\psi_{\boldsymbol{x}+\hat{\mu}}\rangle\neq 0.
\end{equation} 
Combining Eqs.~\eqref{eqn:sp},~\eqref{eqn:a1} and~\eqref{eqn:a2}, we conclude that the transverse components $S_{\boldsymbol{x}}^{\pm}$ of the spin also develop a finite expectation value in the CFM phase. We note that the intersublattice correlations, Eq.~\eqref{eqn:a2}, were indeed found to be nonzero in the gauge mean-field treatment, Eq.~(11) and Eq.~(B1) of Ref.~[\onlinecite{Savary.2012}].  We now demonstrate this with explicit calculation. 

We start our calculation by reviewing some basic properties of the rotor model within the large-$N$ approximation. We consider the rotor Hamiltonian \eqref{eqn:hma}. We follow Ref.~[\onlinecite{Savary.2012}] by relaxing the \emph{local} constraint $|\psi_{\boldsymbol{x}}|=1$ to a global one by adding the following Lagrange multiplier to Eq.~\eqref{eqn:hma}:
\begin{equation}\label{eqn:gc}
\lambda\,[\sum_{\boldsymbol{x}}\psi_{\boldsymbol{x}}^\dagger\psi_{\boldsymbol{x}}-V]
\end{equation}
where $V$ is the number of FCC unit cells. $\psi_{\boldsymbol{x}}=q_{1\boldsymbol{x}}+iq_{2\boldsymbol{x}}$ where $q_{i\boldsymbol{x}}$ ($i=1,2$) are the generalized coordinates. As demonstrated in Ref.~[\onlinecite{Savary.2012}], $Q_{\boldsymbol{x}}=p_{1\boldsymbol{x}}+ip_{2\boldsymbol{x}}$, $p_{i\boldsymbol{x}}$ is the conjugate momentum of $q_{i\boldsymbol{x}}$. Transforming into momentum space, the Hamiltonian \eqref{eqn:hma} is reduced to a collection of non-interacting harmonic oscillators:
\begin{equation}
H=\sum_{\boldsymbol{k}}\sum_{i=1}^2\left[\frac{1}{2}p_{i\boldsymbol{k}}^2+\left(\lambda-j_{\pm}\sum_{\alpha\neq\beta}\cos\frac{k_\alpha}{2}\cos\frac{k_{\beta}}{2}\right)\right]-\lambda V. 
\end{equation}
The dispersion of a single particle excitation is:
\begin{equation}
\omega_{\boldsymbol{k}}=\sqrt{2\lambda-j_{\pm}\sum_{\alpha\neq\beta}\cos\frac{k_\alpha}{2}\cos\frac{k_{\beta}}{2}}
\end{equation}
and the ground state energy per unit cell is:
\begin{equation}\label{app:e0}
E_0(\lambda)=\frac{1}{V}\sum_{\boldsymbol{k}}\omega_{\boldsymbol{k}}-\lambda. 
\end{equation}
$\lambda$ is tuned to satisfy:
\begin{equation}\label{eqn:self-consistent}
\frac{\partial E_0}{\partial \lambda}=0. 
\end{equation}

We perform some simple checks against known results. In the limit of $j_{\pm}\to 0$, Eq.~\eqref{eqn:self-consistent} can be solved exactly with $\lambda=1/2$. We can put this back into Eq.~\eqref{app:e0} to obtain:
\begin{equation*}
E_0=\frac{1}{2}. 
\end{equation*}
For comparision, starting from the original rotor model \eqref{eqn:hma} in the limit $j_{\pm}=0$, the ground state energy is found to be zero. 

In the same limit, the single particle excitation energy is approximately:
\begin{equation}
\omega_{\boldsymbol{k}}\approx 1+O(j_{\pm}),
\end{equation}
which does not agree with the energy cost of value $1/2$ (Eq.~\eqref{eqn:app}) for Hamiltonian \eqref{eqn:hma} as $j_{\pm}\to 0$ limit. 

We now demonstrate that, within framework of the gauge mean-field theory, the CFM phase has ordered XY moment. For this purpose, we focus on demonstrating that the intersublattice correlation $\langle \psi_{\boldsymbol{x}}^\ast \psi_{\boldsymbol{x}+\hat{\mu}}\rangle$ are non-zero. We write Eq. \eqref{eqn:hamigauge} the same way as in Ref.~[\onlinecite{Savary.2012}]:
\begin{widetext}
\begin{subequations}\label{eqn:hamigaugeapp}
\begin{eqnarray}
&&\mathcal{H}=\frac{1}{2}\sum_{\boldsymbol{x}}\hat{Q}_{\boldsymbol{x}}^2-
\sum_{\boldsymbol{x}\in\langle A\rangle}\left[j_{\pm}\sum_{\mu<\nu}(\psi_{\boldsymbol{x}}^\dagger s_{\boldsymbol{x}\mu}^+ s_{\boldsymbol{x}+\hat{\mu}-\hat{\nu},\nu}^-\psi_{\boldsymbol{x}+\hat{\mu}-\hat{\nu}}+\psi_{\boldsymbol{x}+\hat{\mu}}^\dagger s_{\boldsymbol{x}\mu}^-s_{\boldsymbol{x}\nu}^+\psi_{\boldsymbol{x}+\hat{\nu}}+h.c)\right.\label{eqn:ha}\\
&&\left.-j_{z\pm}\sum_{\mu\neq\nu}\left(S_{\boldsymbol{x}\mu}^z(\psi_{\boldsymbol{x}}^\dagger s_{\boldsymbol{x}\nu}^+\psi_{\boldsymbol{x}+\hat{\nu}}+\psi_{\boldsymbol{x}+\hat{\mu}-\hat{\nu}}^\dagger s_{\boldsymbol{x}+\hat{\mu}-\hat{\nu},\nu}^{+}\psi_{\boldsymbol{x}+\hat{\mu}})\mathrm{e}^{i\phi_{\mu\nu}}+h.c\right)\right].\label{eqn:hb}
\end{eqnarray}
\end{subequations}
\end{widetext}
In this section, we only consider $j_{\pm}=0.1$ and $j_{z\pm}=0.1$. This set of parameters would lead to the CFM phase in a self-consistent gauge-mean field calculation, as shown in Fig. 3 of Ref.~[\onlinecite{Savary.2012}]. We use the following ansatz \cite{Savary.2012}:
\begin{equation}
S_{\boldsymbol{x}\mu}^z=\xi_{\mu} \sin\theta,\qquad s_{\boldsymbol{x}{\mu}}^+=\cos\theta
\end{equation}
where $\xi_{\mu}=1,1,-1,-1$ for $\mu=0,1,2,3$. We relax the local constrain $|\psi_{\boldsymbol{x}}|^2=1$ to a global one by adding the Lagrange multiplier \eqref{eqn:gc} for both the A and B sublattices. We represent the two sublattices using subscript $i=1,2$ hereafter. Similarly, $Q_{\boldsymbol{x}}$ becomes (i.e. $Q_{\bm x} \rightarrow \Pi_{\bm x} $) the complex conjugate momentum $\Pi_{\boldsymbol{x}}$ of $\psi_{\boldsymbol{x}}$. In terms of the Bloch modes, the Hamiltonian \eqref{eqn:hamigaugeapp} can be written as:
\begin{equation}\label{eqn:ll}
\mathcal{H}=\sum_{\boldsymbol{k}}\sum_{i,j}\left[\frac{1}{2}\Pi_{i\boldsymbol{k}}^\ast \Pi_{j\boldsymbol{k}}\delta_{ij}+\psi_{i\boldsymbol{k}}^\ast M_{ij}(\boldsymbol{k})\psi_{j\boldsymbol{k}}-2\lambda\right]. 
\end{equation}
The elements of the quadratic kernel $M(\boldsymbol{k})$ are
\begin{subequations}
\begin{eqnarray}
M_{11}(\boldsymbol{k})&=&\lambda-\frac{j_{\pm}}{2}\cos^2\theta\sum_{\alpha\neq\beta}\cos\frac{k_{\alpha}}{2}\cos\frac{k_{\beta}}{2},\nonumber\\
&\equiv&\lambda-\frac{j_{\pm}}{2}\cos^2\theta\rho_{\boldsymbol{k}},\nonumber\\
M_{12}(\boldsymbol{k})&=&-\frac{j_{z\pm}}{4}\sin2\theta\sum_{\mu\neq\nu}\xi_{\mu}\mathrm{e}^{-i\boldsymbol{k}\cdot\hat{\mu}}\mathrm{e}^{i\phi_{\mu\nu}}\equiv -\frac{j_{z\pm}}{4}\sin2\theta h_{\boldsymbol{k}},\nonumber\\
M_{21}(\boldsymbol{k})&=&M_{12}^\ast(\boldsymbol{k}),\nonumber\\
M_{22}(\boldsymbol{k})&=&M_{11}(\boldsymbol{k})\nonumber. 
\end{eqnarray}
\end{subequations}
Diagonalizing $M(\boldsymbol{k})$ by a unitary transformation, \eqref{eqn:ll} becomes a collection of non-interacting harmonic oscillators. Their frequencies are:
\begin{equation}
\omega_{1,2}(\boldsymbol{k})=\sqrt{2\lambda-j_{\pm}\cos^2\theta\rho_{\boldsymbol{k}}\mp \frac{j_{z\pm}}{2}\sin2\theta|h_{\boldsymbol{k}}|}. 
\end{equation}
\begin{figure}[t]
\centering
\includegraphics[width=0.95\columnwidth]{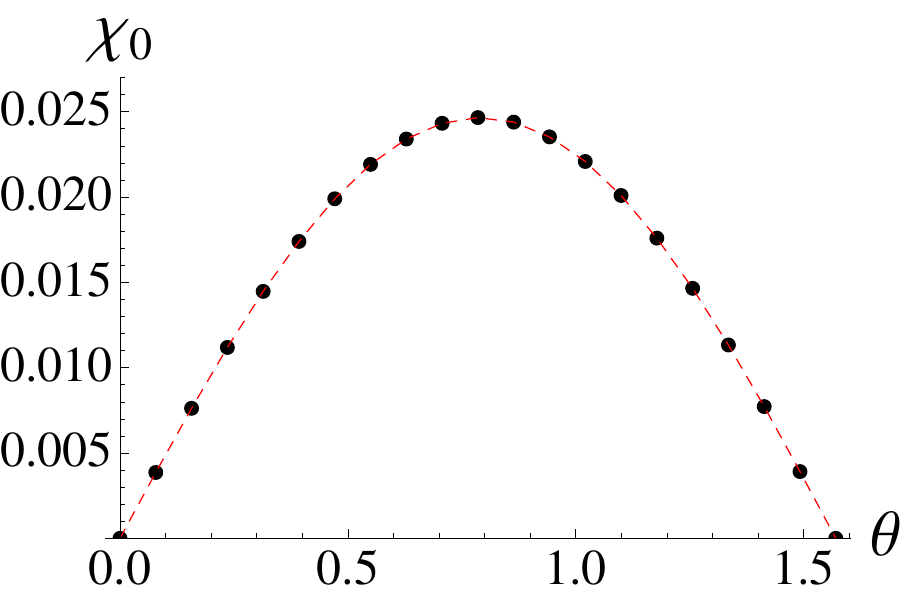}
\caption{$\chi_0\equiv|\langle \psi_{\boldsymbol{x}}^\ast \psi_{\boldsymbol{x}+\hat{0}}\rangle|$ is shown for  $0\le \theta \le \pi/2$. $j_{\pm}=0.1$ and $j_{z\pm}=0.1$.} \label{fig:ex}
\end{figure}

The ground state energy of the spinon sector is:
\begin{equation}
E_0=\sum_{\boldsymbol{k}}\left[\omega_{1}(\boldsymbol{k})+\omega_{2}(\boldsymbol{k})-2\lambda\right]. 
\end{equation}
$\lambda$ is determined by solving the following self-consistent equation:
\begin{equation}
\frac{\partial E_0}{\partial \lambda}=0. 
\end{equation}

The proper procedure is to solve $\theta$ self-consistently \cite{Savary.2012}. However, for our purpose, we only need to show $\langle \psi_{\boldsymbol{x}}^\ast \psi_{\boldsymbol{x}+\hat{\mu}}\rangle$ is finite for $0\le \theta\le \pi/2$ since the self-consistent solution $\theta$ will fall in this range. $\chi_{\mu}\equiv |\langle \psi_{\boldsymbol{x}}^\ast \psi_{\boldsymbol{x}+\hat{\mu}}\rangle|$ is determined by the following equation:
\begin{equation}
\langle \psi_{\boldsymbol{x}}^\ast \psi_{\boldsymbol{x}+\hat{\mu}}\rangle=\frac{1}{N}\sum_{\boldsymbol{k}}\left[\frac{1}{\omega_{1}(\boldsymbol{k})}-\frac{1}{\omega_{2}(\boldsymbol{k})}\right]\mathrm{e}^{-i\phi_{\boldsymbol{k}}}\mathrm{e}^{-i\boldsymbol{k}\cdot\hat{\mu}}
\end{equation}
where $h_{\boldsymbol{k}}\equiv|h_{\boldsymbol{k}}|\mathrm{e}^{i\boldsymbol{k}\cdot\hat{\mu}}$.  $\chi_{\mu}$ is finite for $0\le \theta \le \pi/2$ (Fig.~\ref{fig:ex}). This demonstrate that the $\langle S^{\pm}_{\boldsymbol{x}\mu}\rangle\sim \chi_{\mu}$ is non-zero in general whenever $\langle \psi_{\boldsymbol{x}}^\ast \psi_{\boldsymbol{x}+\hat{\mu}}\rangle$ is nonzero due to $\langle S_{\boldsymbol{x}\mu}^{z}\rangle\neq 0$. 

We note that the CFM self-consistent solution (Eqs.~\eqref{eqn:sp},~\eqref{eqn:a1} and~\eqref{eqn:a2}) exists within the gauge mean-field formalism \cite{Savary.2012}. Within the solution, it costs a finite amount of energy, $\Delta$, to create a pair of deconfined spinons. $\Delta$ vanishes beyond a second phase boundary, which the authors of Ref.~[\onlinecite{Savary.2012}] identified as the phase boundary between the CFM phase and the SFM phase. While the phase has implicit long-range magnetic order, two perspectives can be taken for the nature of such a phase. An ``optimistic'' perspective would assert that the modern definition of an exotic spin phase lies in the long-range entanglement structure of the wave function \cite{Wen.2002}, or ``quantum order''. While the long-range magnetic order in the CFM phase breaks all spin and space symmetries, it is still possible, in principle, that the state has non-trivial quantum order and is thus an exotic phase. One can find a supportive argument for this by following a gauge theory reasonings: the expectation value of the correlator $\langle \psi_{\boldsymbol{x}}^\dagger \psi_{\boldsymbol{x}+\hat{\mu}}\rangle$ can \emph{not} gap out the ``photons'' through the Higgs mechanism since it does not carry any charge. The potential demonstration of the CFM phase being a state with quantum order \emph{and} long-range magnetic order would be a truly remarkable discovery and provides a concrete example of a three-dimensional gapless state with quantum order, which could be realized in real materials. While this is an exciting perspective, there is no concrete numerical or experimental evidence for it yet. 

We thus adopt  a conservative perspective and expect that the implicit magnetic long-range order in all spin components in the CFM phase likely would confine the spinons, contrary to the claim of Ref.~[\onlinecite{Savary.2012}]. From this perspective, the CFM/U$(1)$ liquid phase boundary should be reinterpreted as the phase boundary between some magnetically long-ranged ordered phase, likely the SFM phase, and the U$(1)$ liquid phase within the gauge mean-field formalism. One may speculate that in an ``exact treatment'' of the model other types of singularities could still exist upon crossing the phase boundary between the CFM phase and SFM phase identified in Ref.~[\onlinecite{Savary.2012}], such as a first order jump in the confining string tension between spinons, for example.

\end{appendix}
\bibliography{kramerqsi}
\end{document}